\documentclass[twocolumn, superscriptaddress, aps]{revtex4-1}

\newcommand{\be}{\begin{equation}}
\newcommand{\intf}{\int_{-\infty}^{+\infty}}
\newcommand{\ee}{\end{equation}} 
\newcommand{\bea}{\begin{eqnarray}}
\newcommand{\eea}{\end{eqnarray}}

\newcommand{\ba}{\begin{array}}
\newcommand{\ea}{\end{array}}

\newcommand{\se}{Schr\"{o}dinger equation}
\newcommand{\bl}{\begin{flalign}}
\newcommand{\enl}{\end{flalign}}

\newcommand{\tdse}{time dependent Schr\"{o}dinger equation\ }

\newcommand{\pat}{\frac{d }{d t}}
\newcommand{\ih}{i \hbar}

\usepackage{graphicx}
\usepackage{amssymb}
\usepackage{epstopdf}
\usepackage{amsmath,dsfont}
\usepackage{bm}
\usepackage{braket}
\usepackage{hyperref, xcolor}

\newcommand{\floor}[1]{\lfloor #1 \rfloor}
\newcommand{\eq}[1]{Eq.~\eqref{#1}}
\newcommand{\Eq}[1]{Equation~\eqref{#1}}
\newcommand{\fig}[1]{Fig.~\ref{#1}}
\newcommand{\Fig}[1]{Figure~\ref{#1}}
\newcommand{\stn}[1]{Sec.~\ref{#1}}

\newcommand{\tord}{\mathcal{T}}
\newcommand*{\rom}[1]{\expandafter\@slowromancap\romannumeral #1@}
\renewcommand{\bf}{\mathbf}
\newcommand{\mc}{\mathcal}
\renewcommand{\Re}{\operatorname{Re}}

\newcommand{\tr}{\text{Tr}}

\newcommand{\bs}{\begin{split}}
\newcommand{\es}{\end{split}}

\begin{document}

\title{ Optical absorption properties of laser-dressed matter }
\author{Bing Gu} 
\affiliation{Department of Chemistry, University of Rochester, Rochester NY, 14627}
 \author{Ignacio Franco}
\email{ignacio.franco@rochester.edu}
\affiliation{Department of Chemistry, University of Rochester, Rochester NY, 14627}
\affiliation{Department of Physics, University of Rochester, Rochester NY 14627 }

\begin{abstract}
Characterizing and controlling matter driven far from equilibrium represents a major challenge for science and technology. Here we develop a theory for the optical absorption of electronic materials driven far from equilibrium by resonant and non-resonant lasers. In it, the interaction between matter and the driving light is treated exactly through a Floquet analysis, while the effects of the probing light are captured to first order in perturbation theory.  The resulting equations are reminiscent to those for equilibrium absorption but with the Floquet modes playing the role of the pristine eigenstates. The formalism is employed to characterize the optical properties of a model nanoscale semiconductor dressed by non-resonant light of intermediate intensity (non-perturbative, but non-ionizing). As shown, non-resonant light can reversibly turn this transparent semiconductor into a broadband absorber and open strong absorption/stimulated emission bands at very low frequencies ($\sim$ meV). Further, the absorption spectra of the driven material exhibit periodic features energetically spaced by the photon energy of the driving light that reflect the periodic structure of the Floquet bands. These developments offers a general approach to understand and predict the emergent optical properties of materials dressed by the electric field of light, and catalyze the design of laser-dressed materials with desired optical properties.   
\end{abstract}

\maketitle

\section{Introduction}
In the past century we have made remarkable progress in our ability to design, synthesize, and model novel materials with specific functionalities. Many of the insights and tools that we have developed operate at or near equilibrium where the materials are at the minimum of an appropriate thermodynamic potential. Much less is known, however, about the properties and governing principles of matter driven far from equilibrium \cite{Fleming2008}. In this regime, the effective properties of matter depend on the applied external stimulus and the material response to it. This nonlinear dependency can lead to emergent properties and phenomena that are qualitatively different from those observed near equilibrium, see e.g. Refs. \onlinecite{Paul1990, Grifoni1998, Kohler2005, Rini2007, Fausti2011, Mitrano2016, Mollow1969, Fleischhauer2005, Autler1955, Butikov2001}.

Here we are concerned with the emergent electronic properties of matter driven far from equilibrium by light. In particular, with the ability of these laser-driven materials to absorb light.  We envision a physical situation in which a laser drives matter far from equilibrium, while a second perturbative laser source probes its effective ability to absorb light across the electromagnetic spectrum. To capture and interpret the optical properties of laser-driven materials, here we introduce a generalization of the usual theory of linear optical absorption to this non-equilibrium situation where matter is constantly driven by light. New theoretical tools are needed because, in this regime, there is no stationary reference state and energy is no longer a conserved quantity. Thus, the increase of energy of a system from a given reference state can no longer be used as a criterion for the absorption of photons. In addition, the fluctuation-dissipation theorem \cite{Kubo1963} and Green-Kubo relations \cite{Green1954, Kubo1957a}, that form the basis of the usual theory of linear optical absorption \cite{Mukamel1995, Boyd2008} because they summarize the response of a system near equilibrium to an external weak perturbation, are no longer valid since the Hamiltonian of driven matter is not time-translational invariant. In turn, fully perturbative approaches \cite{Boyd2008} of the response of matter to both driving and probing pulse, while possible, cannot capture the dynamics induced by the driving pulse exactly in all regimes of the laser-matter interaction. 

The theory proposed below overcomes these issues by redefining the absorption properties of driven matter as the rate of transitions induced by the probe photons among the laser-dressed states of the system. From this definition, the optical absorption can be expressed by a non-equilibrium dipole-dipole time-correlation function within first-order perturbation theory in the probing light. In turn, the non-perturbative effects of the driving pulse are captured by introducing a Floquet picture into the analysis and focusing on non-interacting electronic materials for which the equations of motion of the fermionic creation and annihilation operators can be closed exactly. The formalism takes into account the quantum statistics of the pristine material, and the non-equilibrium nature of the laser-dressed matter, and generalizes previous attempts to define the optical properties of laser-dressed matter in various limits  \cite{Mizumoto2006, Qasim2018, Fleischhauer2005, Autler1955, Mollow1969}. Further, it provides a useful starting point for future efforts to capture additional features introduced by electron-electron, electron-nuclear, or other many-body interactions that can contribute to heating and broadening of the spectral features. 

The resulting theory has a structure that is akin to the usual linear absorption theory. However, the photoinduced transitions and transition dipoles that are encountered are between single-particle Floquet eigenstates and not between the pristine eigenstates of the system. To demonstrate its utility, we first apply it to a three-level system under resonant driving. As shown, the formalism naturally recovers the well-known Autler-Townes effect\cite{Autler1955, Cohen-Tannoudji1996} in which a spectral line in the absorption spectrum splits due to near resonance laser driving.

Importantly, the theory also provides the technical means to develop physical insights into the absorption properties of driven-matter, and establish structure-function relations that apply far from equilibrium. In fact, below we use it to explore and interpret the optical properties of a model nanoscale semiconductor dressed by non-resonant lasers. Through Stark effects, non-resonant lasers of intermediate intensity (non-perturbative, but non-ionizing) can strongly modify, in a reversible fashion, the electronic structure of extended and nanoscale materials\cite{Franco2007, Chen2018a, Schiffrin2013, Schultze2013}. As shown below, in addition to exhibiting a redshift in the absorption features reminiscent of the Franz-Keldysh effect \cite{Keldysh1958} and the quantum confined Stark effect \cite{Miller1984},  these laser-dressed materials have novel transient optical properties that are very different from those observed near equilibrium.  In fact, we find that non-resonant light can reversibly turn a transparent semiconductor into a broadband absorber and open strong absorption/stimulated emission bands at very low frequencies ($\sim$ meV).

The structure of this paper is as follows. In \stn{sec:theory}, we introduce the optical absorption theory for laser-dressed matter. The theory relates the absorption properties to the non-equilibrium two-time dipole-dipole correlation function in the interaction picture of the laser-dressed Hamiltonian. Such correlation function is made computationally tractable by adopting a Floquet strategy.  In   \stn{sec:simulation}, the theory is applied to simulate  the non-equilibrium absorption spectrum  of a three-level system under resonant driving and a model nanoscale semiconductor under non-resonant driving.  The simulated non-equilibrium absorption spectra is interpreted in terms of inter- and intra-Brillouin zone transitions between Floquet modes. In \stn{sec:conclude} we summarize our main findings and introduce a qualitative picture for the interpretation of non-equilibrium absorption.

 \section{Theory} \label{sec:theory}
\subsection{Hamiltonian}

We consider the optical properties of a material with Hamiltonian $ H_M$  that is constantly being driven by light. The effective Hamiltonian of this laser-dressed system is
\be
{H}_\text{LD}(t) = {H}_{M} + {H}_{{d}}(t),
\ee
where ${H}_{{d}}(t)= -{\bm \mu}\cdot\bf{E}_{d}(t)$ is the laser-matter interaction in dipole approximation,  $\bf{E}_{d}(t)$ the electric field of the driving light, and $\bm{{\mu}}$ the dipole vector operator. Here and throughout,  boldface denotes vector quantities. The driving laser can be of arbitrary strength and shape, and taken to have periodicity $T$ (angular frequency $\Omega = 2\pi/T$) such that ${H}_{{d}}(t+T)= {H}_{{d}}(t)$. For pulsed excitation sources, this treatment is appropriate when  the envelope of the driving light changes slowly compared to $T$. The optical properties of this laser-dressed material are probed by allowing the material to interact with a weak perturbative probe laser $\bf{E}_{p}(t)$. The total Hamiltonian of the system interacting with both the drive and probe laser is 
\be  { H}(t) = {H}_{{LD}}(t) + {H}_{{p}}(t), \ee
where ${H}_{{p}}= -\bf{\bm \mu}\cdot\bf{E}_{p}(t)$. Because the material is driven out of equilibrium by the driving laser, the equilibrium theory \cite{Mukamel1995} connecting the absorption property and the dipole-dipole correlation function cannot be used here.    In the following, we generalize the definition of absorption spectrum to materials driven far from equilibrium. We adopt the following notation: $\alpha, \beta, \gamma, \delta$ will label single-particle eigenstates of the material Hamiltonian; $\lambda, \eta$ Floquet states; $n$ Fourier components; and $\ket{i}, \ket{f}$ many-electron states.

\subsection{Optical response of non-equilibrium matter} 
For definitiveness, we focus on a system that is initially prepared at time $t_0$ in a particular many-electron state $\ket{i}$ with density matrix $\rho = \ket{i}\bra{i}$. However, the results presented below are general and apply to initial thermal states, and other non-pure states.   To  define the optical absorption for non-equilibrium matter we quantify its response to interaction with a monochromatic probe light at a given frequency $\omega$. Contrary to the equilibrium case, changes in the energy of the system is not a good measure of absorption of light since the energy  of the laser-dressed system is not conserved. The absorption and stimulated emission can be determined by capturing all physical processes that lead to a change of state of the laser-dressed material via interaction with a  photon from the probe laser. The rate at which this happens is given by \cite{Mizumoto2006}
\be I(\omega) = \lim_{t\to \infty} \frac{P(t,\omega)}{ t - t_0}, \label{eq:definition} \ee 
where $P(t,\omega)$ is the probability of a probe photon of frequency $\omega$ to lead to change in the laser-dressed material after an interaction time interval $t-t_0$.

In this analysis, it is useful to decompose the total evolution operator $U(t,t_0)$ of the system into a part $ U_d(t,t_0)= \tord e^{-\frac{i}{\hbar} \int_{t_0}^t  H_\text{LD}(\tau)  \,d\tau} $  that is  due to the driving pulse only, and contributions to the dynamics $ S(t,t_0)$  by the probe light  in the presence of the driving pulse, i.e. 
\be   U(t,t_0) = \tord e^{-(i/\hbar) \int_{t_0}^t {H}(\tau)  \,d\tau} =  U_d(t,t_0)  S(t,t_0), \label{eq:evotot} \ee 
where $\tord$ denotes time-ordering.
To understand the physical processes that contribute to $P(t,\omega)$ we  introduce a transition amplitude $A_{if}$ between two many-body states $\ket{i}$ and $\ket{f}$ of the pristine material of the form,  
\be A_{fi} = \braket{f | U_d^\dagger(t,t_0) U(t,t_0) |i} \equiv \braket{f| S(t,t_0)|i} \label{eq:def0}. \ee 
 The quantity $A_{fi}$ can be interpreted in two complementary but equivalent ways. It can seen as the overlap between the state of the system at time $t$, $ U(t,t_0)\ket{i}$, under the influence of both probe and drive pulses, onto the laser dressed states $ U_d(t,t_0)\ket{f}$. Alternatively,  it can be seen as the projection onto $\ket{f}$ of an initial state that is propagated in a closed time  loop. Such a loop consists of forward propagation from $t_0   \rightarrow t$ with both lasers turned on and then backwards from $t \rightarrow t_0$ with only the driving pulse. This process is akin to the Kelydysh contour used in the Schwinger-Keldysh formalism \cite{Stefanucci2013} and the Loschmidt echo in the study of quantum chaos \cite{Goussev2012}.  

The photon scattering operator (or, equivalently, the evolution operator in the interaction picture of $ H_{\text{LD}}(t)$) $ S(t,t_0)$ satisfies a \se\  $i\hbar \frac{d}{dt} S(t,t_0) = U_d^\dagger(t,t_0) {H}_{ p}(t)  U_d(t,t_0)  S(t,t_0)$ ($S(t_0,t_0) = 1$) and admits a Dyson perturbative expansion. We consider the effect of ${H}_{p}(t)$ to first order in perturbation theory where 
\be  S(t,t_0) =  1 -\frac{i}{\hbar} \int_{t_0}^t  U_{d}^\dag(t_1,t_0)  H_{p}(t_1) U_d(t_1,t_0) dt_1, \ee 
so that 
\be
\begin{split}
A_{fi} &= \braket{f | S(t, t_0) |i} \\ 
& =\Braket{f|  1 -\frac{i}{\hbar} \int_{t_0}^t  U_d^\dag(t_1,t_0)  H_{p}(t_1) U_d(t_1,t_0) dt_1 |i}\label{eq:def}. 
\end{split}
\ee 

There are two different type of processes that contribute to $P(t,\omega)= P^{(1)}(t, \omega) + P^{(2)}(t, \omega)$. Those in which, upon interaction, the probe photon leads to amplitude in laser dressed states $ U_d(t,t_0)\ket{f}$ different from the  laser-dressed initial state $ U_d(t,t_0)\ket{i}$, i.e. 
\begin{widetext}
\be
P^{(1)}(t, \omega) = \sum_{f } |A_{fi}|^2 
=\frac{1}{\hbar^2} \sum_{f} \left| \Braket{f |   \int_{t_0}^t      U^\dagger_d(t_1,t_0) H_{{p}}(t_1) U_d(t_1,t_0) \,dt_1   | i}\right|^2
\ee 
\end{widetext}
where  the set $\{\ket{f}\}$ consist of every many-body state of the complete basis that is orthogonal to $\ket{i}$, i.e. 
\be  1 - \ket{i}\bra{i} = \sum_{f} \ket{f}\bra{f}. \label{eq:p1} \ee 
A second process that leads to absorption/emission of a probe photon, is one in which the probe light interacts with any transient dipole in the  laser-dressed  state $ U_d(t,t_0)\ket{i}$. In this case, the state of the laser-dressed material is not changed but absorption/emission of a probe photon occurs, i.e.
\begin{widetext}
\be 
P^{(2)}(t)  = |A_{ii}|^2 = 1 + 
 \frac{1}{\hbar^2} \left| \Braket{i |   \int_{t_0}^t      U^\dagger_d(t_1,t_0) H_{{p}}(t_1) U_d(t_1,t_0) \,dt_1  | i} \right|^2. 
\label{eq:p2}  
\ee 
\end{widetext}
This contribution is akin to the interaction of an electric field with a permanent dipole in matter. In the non-equilibrium case the dipole can be permanent or be induced by the driving pulse. 

Combining the two processes, and taking into account Eq.~\eqref{eq:p1}, 
\begin{widetext} 
\be 
P(t,\omega) = P^{(1)}(t) + P^{(2)}(t) = \frac{1}{\hbar^2} \iint_{t_0}^t dt_1 dt_2 \braket{i|  U^\dagger_d(t_1, t_0)  H_{{p}}(t_1)  U_d(t_1, t_0) U_d^\dagger(t_2,t_0) H_{{p}}(t_2) U_d(t_2,t_0) |i}  + 1  
  \label{eq:def2}
 \ee
\end{widetext} 
The contribution of the laser-independent term to $P(t,\omega)$ vanishes when calculating the rate in \eq{eq:definition} and will be dropped from this point on. Note that  $P(t,\omega)$ depends quadratically on the field $\bf{E}_p(t)$ as it reflects a first-order photon absorption/emission process. While additional contributions can arise from second order perturbation theory in $ S(t,t_0)$ that also contribute as $|\bf{E}_p(t)|^2$, these contributions vanish in $P(t,\omega)$. Thus, Eq.~\eqref{eq:def2} is consistent up  to second order in  $\bf{E}_p(t)$.

To specify the response, it suffices to consider a monochromatic probe pulse $\bf{E}_{{p}}(t)= \bm {\varepsilon}_{p}\cos(\omega t)$ of frequency $\omega$, amplitude $\varepsilon_p = |\bm{\varepsilon}_{p}|$ and polarization $\bm \varepsilon_{p}/ \varepsilon_{p}$. In this case,  ${H}_{{p}}(t) = -\mu  \varepsilon_{p} \cos(\omega t),$ where $\mu = \bm \mu \cdot \bm  \varepsilon_{p}/\varepsilon_{p}$ is the dipole vector operator projected onto the direction of laser polarization. It then follows that the transition probability can be written in a compact way: 
 \be 
 \begin{split} 
 P(t,\omega) &= \frac{\varepsilon_p^2}{2\hbar^2}  \iint_{t_0}^{t}dt_1 dt_2  C_{\mu \mu}(t_2,t_1) \times \\ & \Re [e^{-i \omega (t_1-t_2)} + e^{-i \omega (t_1+t_2)}].
 \label{eq:def3}
 \end{split} 
  \ee 
 Here  
  \be C_{AB}(t_2,t_1) \equiv \text{Tr}[ \rho  A_I(t_2) B_I(t_1)], \label{eq:twcorrfunc} \ee
is a two-time correlation function  where $ A_I(t), B_I(t)$ are system operators in interaction picture, i.e. $ A_I(t) =  U_d^\dagger(t,t_0) A U_d(t,t_0)$. The final expression for the rate of absorbing/emitting a photon is given by 
\be 
\begin{split} 
I(\omega) &= \lim_{t \rightarrow \infty}  \frac{|\varepsilon_p|^2}{2\hbar^2}\frac{1}{ t - t_0} \iint_{t_0}^{t}dt_1 dt_2  C_{\mu \mu}(t_2,t_1) \times \\ 
&\Re [e^{-i \omega (t_1-t_2)} + e^{-i \omega (t_1+t_2)}].\label{eq:def4} 
\end{split} 
\ee  
When $\bf{E}_d(t)=0$, Eq. \eqref{eq:def4} reduces to the well-known expression $I_{\text{eq}}(\omega) \propto \int C(\tau) e^{-i \omega \tau}d\tau$, where $C(\tau) = \braket{\bf{ \mu}_I \cdot \bf{ \mu}_I(\tau)}$ for equilibrium systems due to the time-translational invariance in this case \cite{zwanzig2001}. 

Equation \eqref{eq:def4} defines the  optical response of matter driven by non-perturbative light. It applies to any material, to pure or mixed initial states with density matrix $\rho$,  and to resonant and non-resonant driving pulses of arbitrary intensity. Nevertheless, numerically these equations are challenging to use directly because they require  propagating the many-body state to long times and back for each frequency and for each pair of interaction times $t_1$ and $t_2$ with the probe field. 

 Below we specialize our considerations to fermionic systems, and show how further progress can be made by invoking Floquet theorem and focusing on effective noninteracting  systems.

\subsection{Optical absorption for laser-dressed electronic materials}

We consider electronic materials that can be described as an effective non-interacting Hamiltonian, as that expected from time-dependent density functional theory \cite{Runge1984, marques2012}. In this case, 
\begin{equation}
 H_{\text{LD}} =\sum_{\alpha\beta} h_{\alpha\beta}(t)  c^{\dagger}_\alpha c_\beta, 
 \label{eq:mH}
\end{equation}
where the operator $c^{\dagger}_\alpha$ (or $c_\alpha$) creates (or annihilates) a fermion in a given single-particle state $ \ket{\alpha}$, and where the time dependence arises from the interaction with the driving laser. 
To calculate $P(t, \omega)$ [\eq{eq:def3}], note that the integrand in this quantity is determined by  the dipole operator $\bf{ \mu}_I(t) = \sum_{\alpha\beta} \bf{ \mu}_{\alpha\beta} c_{\alpha}^\dagger(t) c_{\beta}(t)$ in interaction picture, where
\be
c_{\alpha}(t) =  U_d^\dagger (t,t_0) c_\alpha  U_d (t,t_0).
\ee
is the annihilation operator in interaction representation. 
To incorporate the effect of the driving pulse exactly, it is thus necessary to obtain a closed expression for 
$c_{\alpha}(t).$ 
The equation of motion for the annihilation operator is 
\be i \hbar \frac{d c_{\alpha}(t)}{dt} = [c_{\alpha}(t),  H_I(t)]
\label{eom_c},  \ee
where $ H_I(t) \equiv U_d^\dagger (t,t_0) H_{\text{LD}}(t)  U_d (t,t_0)$ is the Hamiltonian of molecule plus driving pulse $ H_\text{LD}(t)$ in the above interaction picture. For non-interacting Hamiltonians [\eq{eq:mH}], 
\be
 i \hbar \frac{d c_{\alpha}(t)}{dt} = \sum_\beta h_{\alpha \beta}(t)c_\beta(t).
\label{eq:eom}
\ee
These equations can be solved in closed form to give
\be c_\alpha(t) = \sum_{\beta}[ \mc{U}(t,t_0)]_{\alpha \beta}\ c_\beta. \label{eq:cinter}\ee 
where $\mc{U}(t,t_0) \equiv \tord e^{-(i/\hbar)\int_{t_0}^t \mc H(t')\,dt'}$. Here $\mc H$ is the  effective Hamiltonian of each particle in the laser-dressed system in first-quantization with single-particle matrix elements  $h_{\alpha\beta} = \braket{\alpha |\mc{H}|\beta}$. This simplification allows us to introduce Floquet theory at the single-particle level (see Section \ref{sec:floquet}). 
The solution in \eq{eq:cinter} can be verified by inserting it into \eq{eq:eom} and taking into account that 
\be i \hbar \frac{d}{dt}  \mc{U}(t,t_0) =  \mc H(t) \mc{U}(t,t_0),~~~  \mc{U}(t_0,t_0) =  {1}. \label{eq:seu} \ee
The problem of determining $c_\alpha(t)$, and thus $P(t,\omega)$, has now been reduced to the problem of determining the single-particle evolution operator
$ \mc{U}(t,t_0)$. 
 Equations \eqref{eq:seu} and \eqref{eq:cinter} are solved below using Floquet theory.

\subsubsection{Floquet theory for the single-particle evolution operator} \label{sec:floquet}

As the dressed material's Hamiltonian is periodic $ H_\text{LD}(t) =  H_\text{LD}(t+T)$ ($T = 2\pi/\Omega$), so is $\mc H(t) = \mc H(t+T)$.  According to the Floquet theorem \cite{Chu2004, Floquet1883, Kohler2005},  for periodically driven Hamiltonians,   
there exists solutions, so-called Floquet states, to the \se 
\be \ih \frac{d}{dt} \ket{\psi_\lambda(t)} = \mc H(t) \ket{\psi_{\lambda}(t)} \ee 
 of the form 
\be \ket{\psi_\lambda(t)}  = e^{-i \mathcal{E}_\lambda t/\hbar}\ket{\phi_\lambda(t)}, ~~~\ket{\phi_\lambda(t)} = \ket{\phi_\lambda(t+T)}. \label{eq:floquet}\ee 
where the $\ket{\phi_{\lambda}(t)}$ are the so-called Floquet modes and where the  quasi-energies $\mathcal{E}_\lambda$ are uniquely defined in the first Brillouin zone (BZ) $\{ -\hbar\Omega/2 \le   \mathcal{E}_\lambda < \hbar\Omega/2\}$. 
Note that the Floquet states $\{\ket{\psi_{\lambda}(t)}\}$ are single-particle states rather than many-body states.  
While there exists  also many-body Floquet states, in this context, it is much simpler to work at the single-particle level. 

To understand Floquet theorem, consider the eigenstates of the evolution operator after a time $T$ (i.e. from $t$ to $t+T$)
 \be  \mc{U}(T) \ket{\varphi_\lambda} = e^{-i \mathcal{E}_\lambda T/\hbar} \ket{\varphi_\lambda} \label{eq:eigU} \ee 
 with eigenvalues $e^{-i  \mathcal{E}_\lambda T/\hbar}$, where the energies $\mathcal{E}_\lambda$ are defined by the eigenvalue equation. In this section, for simplicity, we take $t_0 = 0 $ and abbreviate $\mc{U}(t) \equiv \mc{U}(t,t_0)$. 
 The Floquet states of the form in Eq.~\eqref{eq:floquet} can be defined  as
 \be \ket{\psi_\lambda(t)} \equiv  \mc{U}(t) \ket{\varphi_\lambda}=  e^{-i\mc{E}_\lambda t/\hbar}\ket{\phi_\lambda(t)}\ee
 where we have defined the Floquet mode 
 \be \ket{\phi_\lambda(t)} = e^{i\mc{E}_\lambda t/\hbar}  \mc{U}(t) \ket{\varphi_\lambda}. \ee 
 To prove Floquet theorem, it suffices to show the Floquet mode satisfies $\ket{\phi_\lambda(t+T)}=\ket{\phi_\lambda(t)}$. This follows because
 \be
 \begin{split}
 \ket{\phi_\lambda(t+T )} & = e^{i\mc{E}_\lambda t/\hbar} e^{i\mc{E}_\lambda T/\hbar}  \mc{U}(t+T, T)   \mc{U}(T) \ket{\varphi_\lambda} \\
 & = e^{i\mc{E}_\lambda t/\hbar}   \mc{U}(t+T, T) \ket{\varphi_\lambda} \\
 & =  e^{i\mc{E}_\lambda t/\hbar}   \mc{U}(t) \ket{\varphi_\lambda} \\ 
 & = \ket{\phi_\lambda(t)} 
\end{split}
 \ee
 where we have used the eigenvalue relation \eq{eq:eigU}. All quasi-energies $\mathcal{E}_\lambda + n\hbar\Omega $  where $n$ is an integer  satisfy the same eigenvalue equation \eqref{eq:eigU} and define the same state. Inserting   Eq. \eqref{eq:floquet} into the \tdse yields  
  \be 
  \left( \mc H(t) - \ih \pat\right)\ket{\phi_\lambda(t)} = \mathcal{E}_\lambda \ket{\phi_\lambda(t)}. \label{eq:fH}
  \ee
  where $\mc H(t) - \ih \pat$ is the Floquet Hamiltonian defined in the extended space-Hilbert space $\otimes$ time. 
For single particle Hamiltonians, \eq{eq:fH} defines the single-particle Floquet modes and their quasienergies.

Since the Floquet modes are periodic function in time, in addition to their usual expansion in a complete single-particle basis in Hilbert space,  they can also be expanded into Fourier components, $\{e^{i n \Omega t}, n \in \mathds{Z} \}$, i.e.  
 \be \ket{\phi_\lambda(t)} = \sum_{n,\beta} F^{(\lambda)}_{n\beta} e^{i n \Omega t} \ket{\beta}.  \label{eq:fmodes}\ee 
 Substituting this expansion into Eq. \eqref{eq:fH}, left multiplying by $\bra{\alpha} e^{-i m \Omega t}$ and averaging over a time period  $\frac{1}{T}\int_0^T dt$ (i.e. taking the inner product in the extended space) 
 gives the eigenvalue equation   
\be
\label{eq:floeig}
\sum_{m ,\beta } \Gamma_{n \alpha; m\beta}F^{(\lambda)}_{m \beta} = \mathcal{E}_\lambda \sum_{m, \beta}  F^{(\lambda)}_{m \beta}.\ee 
Here the matrix elements of the Floquet Hamiltonian are given by
\be \Gamma_{n \alpha; m\beta} = h^{(n-m)}_{\alpha \beta} + n \hbar \Omega \delta_{nm}\delta_{\alpha \beta} 
\label{eq:fham}
\ee 
where $ \mc H^{(n)}$ ($h_{\alpha \beta}^{(n)} \equiv \braket{\alpha | \mc H^{(n)} |\beta}$) is the $n$th Fourier component of the single-particle Hamiltonian of the laser-dressed system,
\be  \mc H^{(n)} \equiv \frac{1}{T} \int_0^{T}\,dt e^{- i n \Omega t}  \mc H(t). \ee 
Equation \eqref{eq:floeig} is a generalized eigenvalue problem defined in a composite basis $\{\ket{\alpha n } \equiv \ket{\alpha} \otimes e^{i n \Omega t} \| \alpha \in [1,N], n \in \mathds{Z} \}$, where $\ket{\alpha}$ is any complete basis of the Hilbert space. It can be solved to determine the Floquet states and energies. 

Once the Floquet modes $\{\ket{\phi_\lambda(t)} \}$ are determined [\eq{eq:fmodes}]
so will be the propagator $ \mc{U}(t, t_0)$ 
\be  
\mc{U}(t,t_0)  = \sum_\lambda e^{-i \mc{E}_\lambda (t-t_0)/\hbar} \ket{\phi_\lambda(t)} \bra{\phi_\lambda(t_0)}.  \label{eq:bsln}
\ee 
where we have made the initial time dependence explicit. 
Equation \eqref{eq:bsln}  satisfies the \se\ in Eq.~\eqref{eq:seu}, as required.
Substituting the above into Eq.~\eqref{eq:cinter}, leads to the solution for $c_{\alpha}(t)$   
\be c_{\alpha}(t) = \sum_\lambda  e^{-i \mc{E}_\lambda (t-t_0)/\hbar}\sum_{n \beta} F_{n\alpha}^{(\lambda)}e^{i n \Omega t}\braket{\phi_\lambda(t_0)|\beta} c_\beta \label{eq:sol} \ee 
Equations \eqref{eq:bsln} and~\eqref{eq:sol} can now be used to compute correlation functions and the spectrum as described below.

\subsubsection{Correlation function in Floquet theory}

To calculate the two-time dipole-dipole correlation function [\eq{eq:twcorrfunc}],  the dipole operator in the  interaction picture of $H_{\text{LD}}(t)$ is required. To obtain it, one can directly insert the creation and annihilation operators in \eq{eq:sol}  into the dipole operator $\mu_I(t) = U_d^\dagger (t,t_0)\sum_r \tilde{\mu}(r)U_d (t,t_0) = \sum_{\alpha \beta} \mu_{\alpha \beta} c_\alpha^\dag(t) c_\beta(t)$ where $r$ runs over particles and $\mu_{\alpha\beta} = \Braket{\alpha|\tilde{\mu}|\beta}$.  However, here it is simpler, and equivalent, to first take   
the  dipole operator for a single particle $\tilde{\mu}$ and compute its form in the interaction picture of $\mc{H}(t)$, and then use it to construct a second quantized form for $\mu_I(t)$. The equivalence is due to the fact that,   for effective non-interacting electronic systems,  the single-particle operator $\mu_I(t)$ in the interaction picture of $ H_{\text{LD}}(t)$ can be obtained by computing its counterpart in the first-quantized form $ \mc{U}^\dag(t,t_0) \tilde{\mu}  \mc{U}(t,t_0)$ followed by a second-quantization step.   As shown below, this treatment leads to transition dipoles between Floquet modes, offering a compact expression for the final non-equilibrium absorption.  Using this fact and inserting \eq{eq:bsln} for the single-particle evolution operator, it follows that

\be 
\begin{split} 
\bf{\mu}_I(t)  & = \sum_{\gamma,\delta} \Braket{\gamma|\mc{U}^\dag(t,t_0) \tilde{\mu} \mc{U}(t,t_0) |\delta}c_\gamma^\dag c_\delta  \\ 
&= \sum_{\lambda' \lambda} \sum_{\gamma \delta}  \bf{\mu}_{\lambda'\lambda}(t) e^{i \mc{E}_{\lambda'\lambda}(t-t_0)/\hbar} \braket{\gamma |\phi_{\lambda'}^0}\braket{\phi_\lambda^0|\delta} c_\gamma^\dag c_\delta
\label{eq:dipole0}
\end{split} 
\ee 
where 
$ \mu_{\lambda' \lambda}(t) = \braket{\phi_{\lambda'}(t)|\tilde{\mu} |\phi_\lambda(t)} $ is the time-dependent transition dipole between Floquet modes, $\ket{\phi_\lambda^0} \equiv \ket{\phi_{\lambda}(t_0)}$ and $\mc{E}_{\lambda'\lambda} = \mc{E}_{\lambda'} - \mc{E}_\lambda$. 

Because the Floquet modes are periodic, so is the dipole matrix $\bf{\mu}_{\lambda'\lambda}(t) = \bf{\mu}_{\lambda' \lambda}(t+T)$ such that it admits a Fourier expansion 
\be \bf{\mu}_{\lambda'\lambda}(t) =  \sum_{n = -\infty}^\infty \bf{\mu}_{\lambda'\lambda}^{(n)}e^{in\Omega t} 
\label{eq:mufourier}
\ee
with the expansion coefficients 
\be {\mu}_{\lambda'\lambda}^{(n)}(t) = \frac{1}{T}\int_0^T \bf{\mu}_{\lambda'\lambda}(t)e^{-in\Omega t}\,dt \label{eq:eff_dipole} .\ee   
Inserting this expansion into \eq{eq:dipole0} yields   
\be \mu_I(t) = \sum_{\lambda',\lambda, \gamma,\delta} \sum_n D^n_{\lambda' \lambda \gamma \delta} e^{i \mc{E}_{\lambda'\lambda}(t-t_0)/\hbar + i n\Omega t} c_\gamma^\dag c_\delta 
\label{eq:dipole}
\ee  
where 
\be D^n_{\lambda' \lambda \gamma \delta} = { \mu}_{\lambda' \lambda}^{(n)} \Braket{\gamma|\phi_{\lambda'}^0} \Braket{\phi_{\lambda}^0|\delta} . \label{eq:Dcoef} \ee

The correlation function can then be obtained by inserting \eq{eq:dipole} into  \eq{eq:twcorrfunc}
\begin{widetext} 
 \be 
 C_{\mu\mu}(\bar{t},\tau)  =  \sum_{n,n'}\sum_{\lambda, \lambda',\eta,\eta'} \sum_{\gamma \delta \gamma' \delta'} D^n_{ \lambda' \lambda \gamma \delta}  D^{n'}_{\eta' \eta \gamma' \delta'}  e^{i (\mc{E}_{\eta'\eta}  + \mc{E}_{\lambda'\lambda}) (\bar{t}-t_0)/\hbar + i(n'+n) \Omega   \bar{t} }  
  e^{i ((\mc{E}_{\eta'\eta} - \mc{E}_{\lambda'\lambda})/\hbar + (n'-n) \Omega) \tau/2} \braket{ c^\dagger_\gamma  c_\delta  c^\dagger_{\gamma'}  c_{\delta'} } \label{eq:corr0},
\ee 
\end{widetext} 
 where, for future convenience, we have transformed the two time arguments into a center of mass $\bar{t} = \frac{t_1+t_2}{2}$ and a relative time variable $\tau = t_2-t_1$.
For a system initially prepared in a statistical mixture of single Slater determinants, the term $\braket{ c^\dagger_\gamma  c_\delta  c^\dagger_{\gamma'}  c_{\delta'}}$  entering into the correlation function [\eq{eq:corr0}] can be computed as follows.  This term does not vanish in two different cases, $\gamma = \delta, \gamma' = \delta'$ and $\gamma = \delta', \delta = \gamma'$,  which gives 
\be \Lambda_{\gamma \delta \gamma' \delta'} \equiv \braket{c^\dagger_\gamma c_\delta c^\dagger_{\gamma'} c_{\delta'} } =  \delta_{\gamma \delta} \delta_{\gamma' \delta'} \bar{n}_\gamma \bar{n}_{\gamma'} + \delta_{\gamma'\delta}\delta_{\gamma \delta'} \bar{n}_\gamma (1- \bar{n}_{\gamma'}) \label{eq:gamma} \ee 
where $n_\gamma \equiv c_\gamma^\dagger c_\gamma $ is the number operator and $\bar{n}_{\gamma} = \tr\{ \rho n_\gamma\}$ the initial  distribution function of the single-particle energy eigenstates. For thermal initial states  $\bar{n}_{\gamma}$ corresponds to the Fermi-Dirac distribution.

\subsubsection{Time integration and final expressions}

In the center of mass and relative time variables the rate of absorption/emission [\eq{eq:def4}] is given by
\be
\begin{split} 
 I(\omega) =& \lim_{t \rightarrow \infty}  \frac{\varepsilon_p^2}{2\hbar^2}\frac{1}{ t-t_0} \iint_{t_0}^{t}d t_1 dt_2  C_{\mu \mu}(\bar{t},\tau) \times \\  &\Re [e^{-i \omega \tau} + e^{-i 2\omega \bar{t}}]. 
\end{split} 
\ee  
We take the preparation time of the system to be in the remote past, such that $t_0 \rightarrow -\infty$. In this limit,  the two-time integral in \eq{eq:def4} reduces to Fourier transforms, i.e. 
\be 
\begin{split} 
I(\omega) = &\lim_{t \rightarrow \infty}  \frac{\varepsilon_p^2}{2\hbar^2}\frac{1}{ t-t_0} \iint_{-\infty}^{t}d \bar t d\tau  C_{\mu \mu}(\bar{t},\tau) \\  &\Re [e^{-i \omega \tau} + e^{-i 2\omega \bar{t}}].
\end{split} 
\label{eq:tmp}
\ee 
The second complex exponential term that depend on $\bar{t}$ in \eq{eq:tmp} does not contribute to  $I(\omega)$, see Appendix A for details. 
It suffices then to focus on the $e^{-i\omega \tau}$ term, i.e.
\be
I(\omega) =   \frac{\varepsilon_p^2}{4\hbar^2}  \lim_{t \rightarrow +\infty} \frac{1}{ t - t_0}\iint_{-\infty}^{t} C_{\mu \mu}(\bar{t},\tau) (e^{-i \omega \tau} + \text{c.c.})\, d\bar{t}d\tau \\ 
\ee
Inserting \eq{eq:corr0} into the above equation, one notices that the integration with respect to  $ \bar{t}$ gives oscillatory contributions whose contribution to $I(\omega)$ vanishes at $ t \to +\infty$ except when the oscillatory factor is zero. In that case, the integration leads to a $(t-t_0)$ term that cancels the $1/ (t-t_0)$ in the expression for $I(\omega)$. This happens when $\mc{E}_{\eta'\eta} + \mc{E}_{\lambda'\lambda} = 0$ and $n+n' = 0$. The former condition implies that either $\eta' = \lambda, \eta = \lambda'$ or $\eta' = \eta, \lambda' = \lambda$.
Taking this into account, the  absorption spectrum can be written as 
\be
\begin{split}
I(\omega)  = &  \frac{\varepsilon_p^2}{4h} \sum_{\gamma \delta \gamma' \delta'} \sum_{\lambda,\lambda'}     \sum_{n} \bigg[ D^n_{\lambda \lambda \gamma\delta}D^{-n}_{\lambda' \lambda' \gamma' \delta'} \delta(n \hbar \Omega - \hbar\omega) \\ 
& +   D^{-n}_{\lambda \lambda' \gamma\delta}D^{n}_{\lambda'\lambda \gamma' \delta'} \delta(\mathcal{E}_{\lambda'\lambda} + n\hbar \Omega - \hbar\omega) \bigg] \Lambda_{\gamma \delta \gamma' \delta'}  \\
&+ \left( \omega \leftrightarrow -\omega\right) 
\label{eq:final} 
\end{split}
\ee
where the last term corresponds to the same expression but replacing  $\omega$ with $-\omega$, and where we have taken into account of the integral representation of the delta function $\delta(\omega) = \frac{1}{2\pi}\intf e^{i\omega t}\,dt$ and $\delta(\hbar \omega) = \delta(\omega)/\hbar$.

The quantity $I(\omega)$ measures the rate of change induced by the probe photons on the laser dressed material. However,  it does not tell us whether the change is due to absorption or stimulated emission processes. We  identify 
the first two terms in \eq{eq:final}  as optical absorption because $\omega>0$ and the delta functions in \eq{eq:final} will only be non-zero when the energy difference between the Floquet states involved $\mc{E}_{\lambda'\lambda} + n \hbar  \Omega$ is positive, leading to absorption of photons from the probe field. By contrast, the $-\omega$ term corresponds to stimulated emission. The net absorption of probe photons by the laser-dressed material would correspond to the difference between these two contributions:
\be 
\begin{split} 
	A(\omega) & =   \frac{|\varepsilon_p|^2}{4h} \sum_{\lambda,\lambda'}\sum_{\gamma \gamma'} \sum_n \left(    D^{-n}_{\lambda \lambda' \gamma\gamma'}D^{n}_{\lambda'\lambda \gamma' \gamma}\right) \\ 
	& \times 
	\bigg[\delta(\mathcal{E}_{\lambda'\lambda} + n \hbar \Omega - \hbar\omega) - \delta(\mathcal{E}_{\lambda'\lambda} + n\hbar  \Omega + \hbar\omega)\bigg] \\ 
	&\times   \bar{n}_\gamma(1-\bar{n}_{\gamma'})
\end{split}
\label{eq:finalA2}
\ee
where we have taken into account that the $\delta(n \hbar \Omega - \hbar \omega)$  and $\bar{n}_\gamma \bar{n}_{\gamma'}$ contribution exactly cancel.
Inserting \eq{eq:Dcoef} into \eq{eq:finalA2} and introducing 
\be P_{\lambda \lambda'} \equiv \sum_{\gamma\gamma'} |\Braket{\phi_\lambda^0|\gamma}|^2| \Braket{\phi_{\lambda'}^0|{\gamma'}}|^2 \bar{n}_\gamma (1-\bar{n}_{\gamma'}), \label{eq:pop}
\ee
which acts as an effective population factor between the Floquet modes yields the final expression of the absorption spectrum for laser-dressed matter 
\be 	
\begin{split} 
A(\omega)  & =   \frac{|\varepsilon_p|^2}{4h} \sum_{\lambda,\lambda'}\sum_n   | \mu^{(n)}_{\lambda' \lambda}|^2 P_{\lambda \lambda'} \times \\
& (\delta(\mathcal{E}_{\lambda'\lambda} + n \hbar \Omega - \hbar \omega ) - \delta(\mathcal{E}_{\lambda' \lambda} + n \hbar \Omega + \hbar \omega)) 
\end{split} 
\label{eq:final3}
\ee 
where we have taken into account that $\mu^{(-n)}_{\lambda \lambda'} = \mu^{(n)*}_{\lambda' \lambda}$.

  \Eq{eq:final3} offers a clear structure for the interpretation of non-equilibrium absorption, that is  analogous to the one encountered in equilibrium absorption theory. The Floquet modes play the role of system eigenstates and the effective population factor $P_{\lambda \lambda'}$ characterize the probability that  $\ket{\phi_\lambda}$ is occupied while the state $\ket{\phi_{\lambda'}}$ is unoccupied. The first term captures absorption when the frequency of the probing light is at resonance with a transition frequency between two Floquet modes $\mc{E}_{\lambda' \lambda} + n\hbar \Omega$. In turn, the second term is stimulated emission. The states must be connected by a non-zero transition dipole $\mu_{\lambda \lambda'}^{(n)}$ for a transition to occur.

An additional feature that arises from the  time-dependence of Floquet states is that  the effective dipole operator $\mu_{\lambda \lambda'}^{(n)}$  has an extra index $n$, originating from the periodicity of the Floquet states. This extra index  can be understood as the indicator for intra- or inter-BZ transitions. When $n=0$, it indicates that the transitions are inside the same BZ. In turn, when $n \ne 0$ the transitions happen between different BZs and $n$ indicates the number of BZs that seperates the two Floquet states. This transition is analogous to the  umklapp process in solids where the crystal momentum is changed into another BZ as a result of a scattering process.  The probability for different number of BZs to be involved depends on the details of the system, the strength and frequency of the driving laser. Note that while for equilibrium absorption $\mu_{\alpha \beta} =\mu_{ \beta\alpha}^\star$, for non-equilibrium absorption $\mu_{\lambda \lambda'}^{(n)}\ne \mu_{\lambda' \lambda}^{(n)}$ except for $n=0$.  

\Eq{eq:final3}  shows that one can naturally interpret nonequilibrium absorption as optical transitions among Floquet states. 
 Besides atoms, molecules and nanoscale systems, \eq{eq:final3} can also be applied to solids if the single-particle states are taken as Bloch states. In that case, it is interesting to contrast \eq{eq:final3} to previous efforts to develop theories of laser-dressed semiconductors~\cite{Mizumoto2006, Combescot1989}. \Eq{eq:final3}  generalizes the results in \cite{Mizumoto2006} by providing a physically transparent derivation of the non-equilibrium optical absorption and clarifying its basic structure, incorporating the effects of quantum statistics and, importantly, by recognizing the role of stimulated emission processes in $P(t,\omega)$. 

By adopting a Floquet strategy, we have been able to reduce the dynamic problem of optical absorption/stimulated emission of laser dressed matter to a static problem that requires sums over  Floquet states and single-particle energy eigenstates. These states can be obtained via simple diagonalization techniques. To calculate $A(\omega)$ it is necessary to: (i) Diagonalize the material Hamiltonian $ H_M$ to obtain the single-particle energy eigenstates $\{\ket{\alpha}\}$ and express the dipole operator in this basis; (ii) Construct the Floquet Hamiltonian matrix \eq{eq:fham} and solve \eq{eq:floeig} by diagonalization to obtain  the quasi-energies $\{\mc{E}_\lambda\}$ and the expansion coefficients of Floquet modes $\{F_{n\beta}\}$ in the $\ket{\beta n}$ basis. In practice, to solve these equations the Floquet Hamiltonian matrix needs to be truncated. Results need to be checked for convergence on the number of Fourier components; (iii) Calculate the effective dipole using \eq{eq:eff_dipole} and population factor following the definition in \eq{eq:pop} and use them to compute the absorption spectrum based on Eq. \eqref{eq:final3}.  From a numerical perspective, the second step is most challenging  because it involves a diagonalization of the Floquet matrix whose size scales as $\mc{O}(N_bN_F)$ where $N_F$ is the number of Fourier components and $N_b$ is the number of single-particle orbitals of the system. This matrix grows quickly for realistic systems under non-resonant or strong driving.

\section{Applications of the theory and interpretation of the non-equilibrium spectra} \label{sec:simulation}

Using \eq{eq:final3} we are now in  a position to quantify and interpret the optical properties of laser-dressed matter. The validity of the theory is demonstrated by using it to  recover the well-known Autler-Townes effect of laser-dressed few-level systems. The utility of the approach, by using it to explore the optical properties of nanoscale semiconductors driven by non-resonant light. As shown, non-perturbative reversible driving with non-resonant light  can significantly distort the absorption spectrum leading to a new laser-dressed material with  spectral features that have no equilibrium counterpart. A qualitative scheme to interpret non-equilibrium absorption in the laser-dressed picture is developed and used to assign spectral features in both cases.

\subsection{Resonantly driven three-level system}
\label{stn:AT}

\begin{figure*}[htbp]
	\centering
	\includegraphics[width=0.7\textwidth]{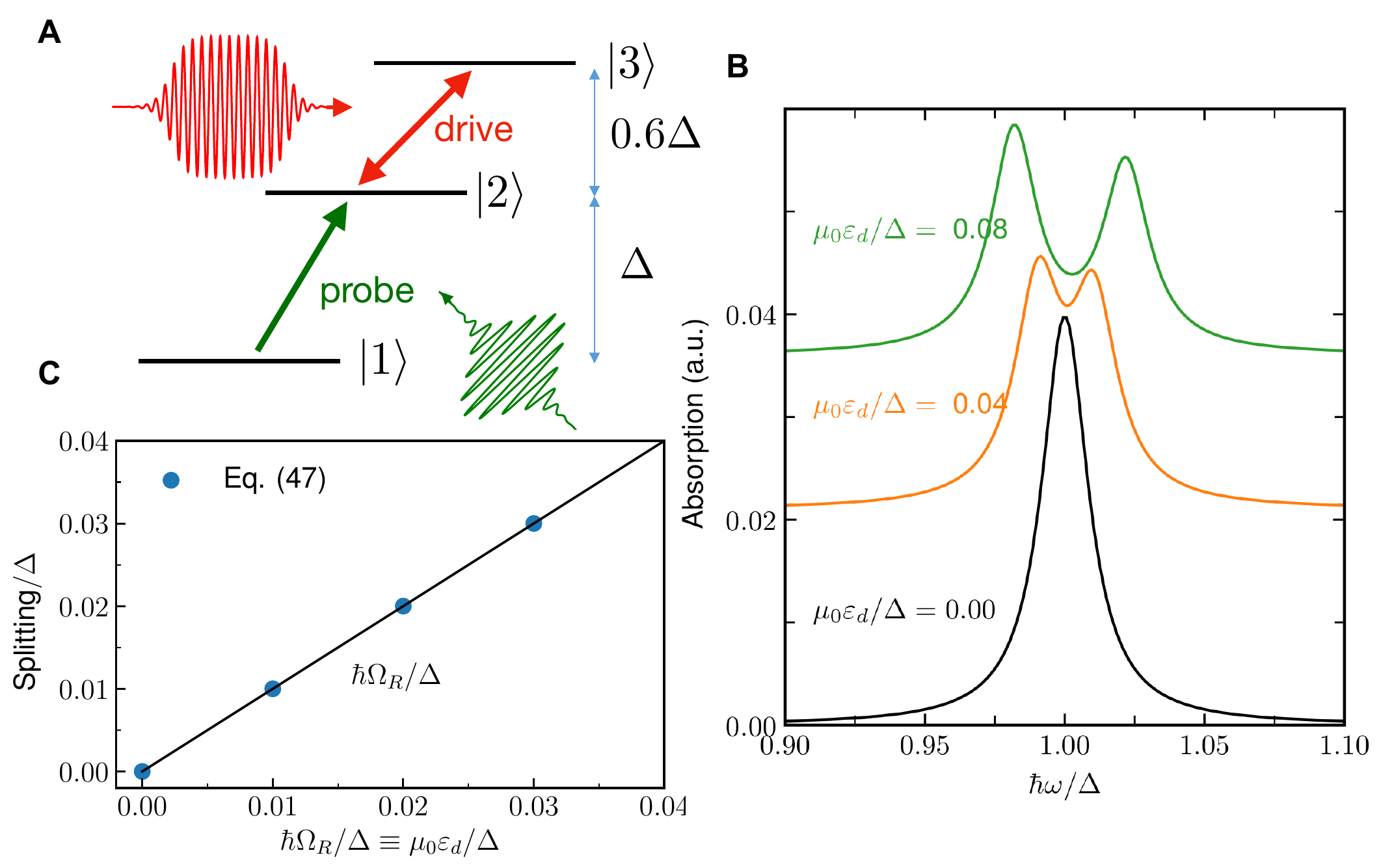}
	\caption{Simulation of the Autler-Townes splitting using \eq{eq:final3}. (A)  Energy diagram for a three-level system. The driving pulse  resonantly drives the $\ket{2}\to\ket{3}$ transition, while the probe light measures the optical absorption. Here $\mu_0=\bra{1}\mu\ket{2}=\bra{2}\mu\ket{3}$ are the non-zero transition dipoles. 
(B) The absorption spectrum computed using  \eq{eq:final3}  (broadened with a Lorentzian function of width $0.01\Delta$) under different driving amplitudes $\varepsilon_d$ naturally exhibits the Autler-Townes splitting. (C) The AT splitting yielded by \eq{eq:final3} is $\hbar\Omega_\text{R}$ where $\Omega_\text{R} \equiv \mu_0\varepsilon_d/\hbar$ is the Rabi frequency in quantitative agreement with theoretical predictions using a different method \cite{Cohen-Tannoudji1996} and experiments~\cite{Xu2007}.}
	\label{fig:resonant} 
\end{figure*}

\begin{figure*}[htbp]
	\centering
	\includegraphics[width=0.7\textwidth]{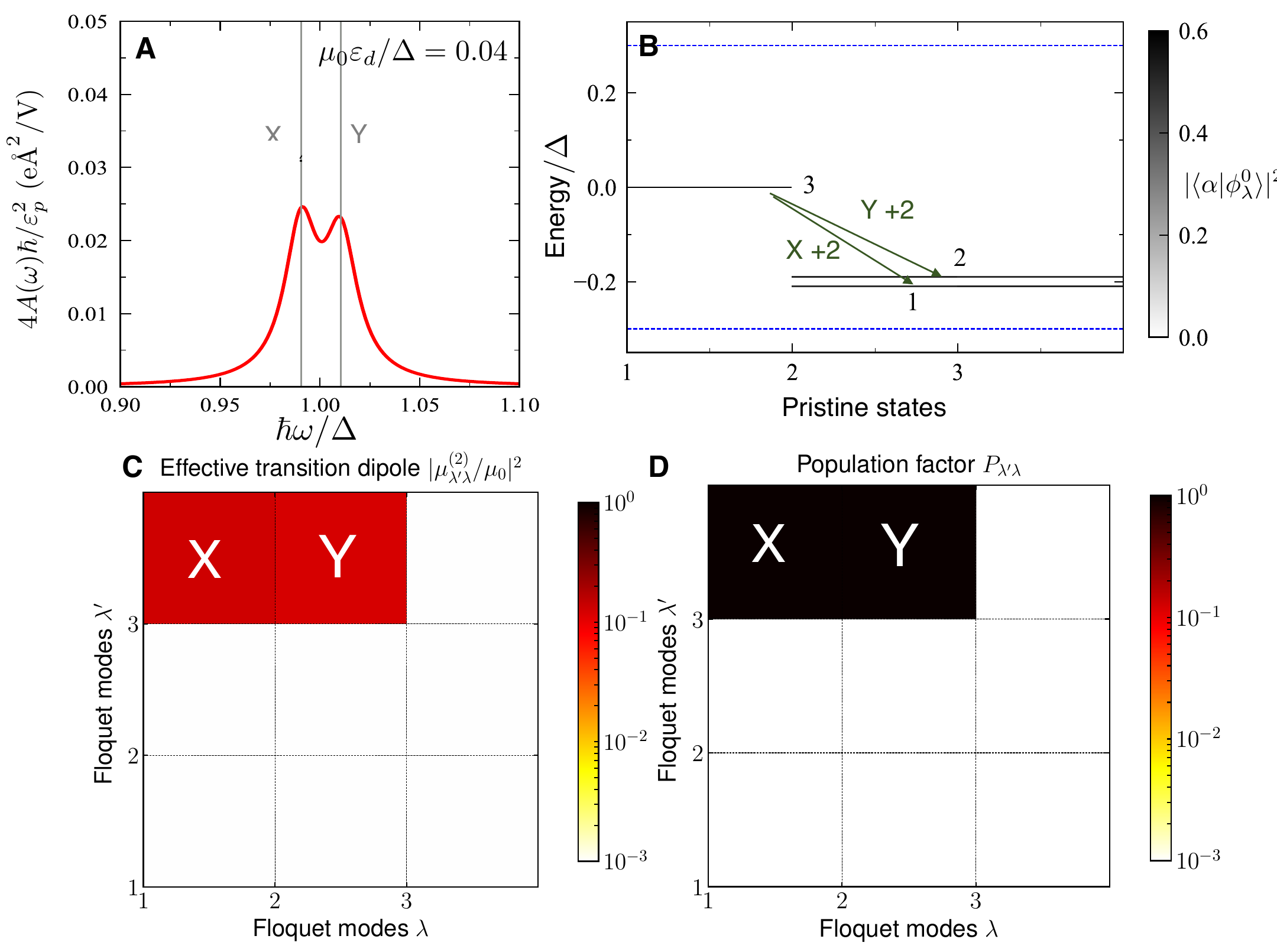}
	\caption{Interpretation of the non-equilibrium absorption spectra  using \eq{eq:final3} of the three-level system in \fig{fig:resonant} initially prepared in $\ket{1}$ dressed by  light resonant with the $\ket{2}\to\ket{3}$ transition with $\mu_0\varepsilon_d/\Delta = 0.04$. (A) The absorption spectrum has  two main transitions labeled X and Y.  (B) Overlap between the Floquet modes at  times $t_0 + nT$ and the pristine states $|\braket{\alpha|\phi_{\lambda}(t_0)}|^2$ in the first BZ. The Floquet modes are ordered by the quasienergy in the first BZ. The optical transitions X and Y  are indicated by green arrows with the number of BZs that separates them $+n$. (C) Effective dipole between Floquet modes two BZs away. Other non-zero transition dipoles are not allowed by the population factor.  (D) Effective population factor $P_{\lambda' \lambda}$ $(\lambda'\to\lambda)$ between Floquet modes. The elements in panels C and D corresponding to the X, Y transitions  are marked.} 
	\label{fig:resonant_interpret} 
\end{figure*}

Autler and Townes \cite{Autler1955} showed that  an optical transition between two levels in a few level system can be split into a doublet  when one of the two levels involved in the transition is coupled to a third one by a strong resonant light, a phenomenon that is also referred as the dynamic Stark splitting. The Autler-Townes (AT) effect has been observed in the absorption spectra of  atoms  \cite{Autler1955}, small molecules \cite{Garcia-Fernandez2005}, superconducting Josephson junction \cite{Sillanpaa2009} and quantum dots \cite{Xu2007} dressed by resonant lasers. 

To demonstrate that \eq{eq:final3} recovers the AT effect,  we computed the non-equilibrium absorption spectra in the three-level system (with states $\ket{1}$, $\ket{2}$ and $\ket{3}$) shown in \fig{fig:resonant}A. In the computations, the system is driven by a laser that is resonant with the $\ket{2} \rightarrow \ket{3}$ transition. The resulting absorption spectra is shown in \fig{fig:resonant}B. As can be seen, the absorption spectrum clearly exhibits the Autler-Townes splitting, and the slope of the observed linear increase in the splitting with $\varepsilon_d$ is in quantitative agreement with previous theoretical and experimental observations (\fig{fig:resonant}C) \cite{Xu2007}.

The interpretation of the AT effect is a well developed subject~\cite{Cohen-Tannoudji1996, Fleischhauer2005}. In the laser-dressed picture it can be understood through resonances induced by resonant driving between Floquet states, and their subsequent Rabi splitting. It is instructive to interpret this phenomenon through \eq{eq:final3}. For definitiveness, consider the   $\mu_0\varepsilon_d/\Delta = 0.04$ case.  Figure \ref{fig:resonant_interpret}B shows the quasienergies in the first BZ and overlap between Floquet modes at times $t_0+nT$ and the pristine states. As can be seen, the Floquet modes $\ket{\phi_1}, \ket{\phi_2}$ are a linear combination of the two pristine states $\ket{2}$ and $\ket{3}$, that are under resonant driving while the Floquet mode $\ket{\phi_3}$ is just the pristine state $\ket{1}$. The two transitions, labeled by X, Y in panel A, are transitions between Floquet modes $\ket{\phi_3} \rightarrow \ket{\phi_1}, \ket{\phi_2}$ separated by two BZs, respectively.  The corresponding effective transition dipole and population factor for these two transitions are marked in \fig{fig:resonant_interpret}C-D. Clearly, these two transitions are allowed by population and transition dipoles.  Other intra and interband transitions with non-zero transition dipoles are not allowed by the population factor.

\subsection{Non-resonantly driven tight-binding nanostructure}
\label{stn:nonresonant}

We now focus on the optical properties of a generic two-band semiconducting nanostructure driven by non-resonant light of intermediate intensity. Through Stark effects, non-resonant light can dramatically distort the electronic structure of nanostructures and extended systems creating a laser-dressed material with effective electronic properties that can be very different from those observed near equilibrium. Below we clarify the optical properties of such laser-dressed materials in the context of a minimal one-dimensional  tight-binding model. We focus on the reversible regime of the laser-matter interaction where the net absorption of photons by matter from the non-resonant driving pulse is suppressed.

The tight-binding Hamiltonian of a one-dimensional two-band semiconducting nanoparticle  with $K$ unit cells is 
\be  
\begin{split} 
H_M &=  \sum_{k=1}^{K}   (\epsilon_1 c_{2k-1}^\dagger c_{2k-1} + \epsilon_2 c_{2k}^\dagger c_{2k}) \\ 
&- \sum_{k=1}^{K}  t_\alpha \left(c_{2k-1}^\dagger c_{2k} + \text{h.c.} \right) - \sum_{k=1}^{K-1} t_\beta
\left( c_{2k}^\dagger c_{2k+1} + \text{h.c.}\right)  
\end{split} 
\label{eq:semi}
\ee
where $c_k^\dagger$ creates a fermion on site $k$ ($\ket{k} =c_k^\dagger\ket{0}$, where $\ket{0}$ is the vaccum state),  and where h.c. stands for hermitian conjugate. Each unit cell consists of two sites with onsite energies $\epsilon_1$ and  $\epsilon_2$ ($\epsilon_1 = -\epsilon_2 =$ 1.6 \text{eV}) in nearest-neighbor coupling with intracell hopping parameter $t_\alpha = 0.7$ eV  and intercell $t_\beta = 1.0$ eV. The lattice constant is taken to be $a =$ 3.2 \AA~ and the two sites in each cell to be separated by a distance $b=0.0$ \AA. These parameters are chosen to resemble the electronic structure of ZnO.  

The nanostructure is dressed in dipole approximation by a non-resonant monochromatic laser field  with electric field amplitude $E_d(t) =\mc{\varepsilon}_d \cos(\Omega t)$, and probed with a laser of  amplitude $E_p(t) =\mc{\varepsilon}_p \cos(\omega t)$.  Both probe and drive are taken to have their polarization along the length of the  nanostructure. At initial time $t_0$ the system is chosen to be in the ground zero-temperature state. While large system sizes can in principle be considered,  below we focus on the $K=6$ case such that a detailed analysis of all transitions visible in the absorption spectrum is tractable. 

We focus on the regime where the frequency of the driving  field is much smaller than the band gap, $ \hbar \Omega = 0.38$ eV $\ll E_g= 3.31$ eV such that Stark effects and not near-resonance multi-photon absorption effects dominate the dynamics. In this regime, the laser-matter interaction is reversible in the sense that, for pulsed driving, after the driving pulse is turned off the system will return to its initial unexcited state. We verify that we are in this regime by explicitly solving the time-dependent Schr\"odinger equation for the nanostructure under the influence of 200 fs Gaussian pulses with the maximum field amplitude and frequency of $E_d$, and ensuring that after the pulse there is no net excitation of the chain.

\begin{figure*}[htbp]
	\centering
	\includegraphics[width=0.9\textwidth]{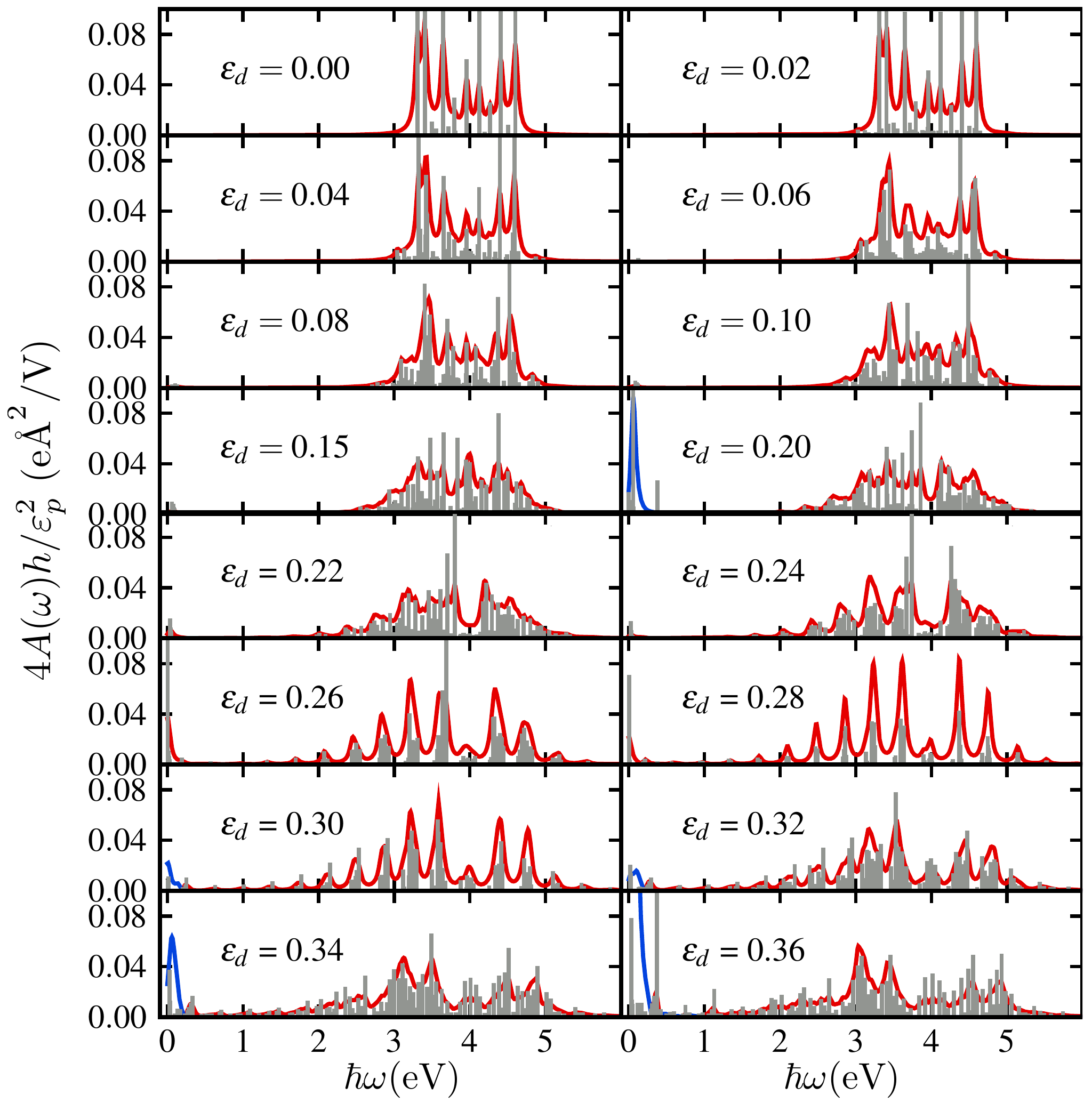}
	\caption{ Linear optical absorption spectrum for a semiconducting nanoparticle with Hamiltonian \eq{eq:semi}  dressed with a continuous wave laser of varying amplitude $\mc{\varepsilon}_d$ (in V/\AA).   Red lines indicate net absorption, blue ones indicate stimulated emission. The gray peaks signal the  frequency and amplitude of each absorption transition between Floquet modes. Peaks are broadened by a Lorentzian function of width $\sigma = 0.04$ eV.   }
	\label{fig:ZnO}
\end{figure*}

\Fig{fig:ZnO} shows the non-equilibrium  absorption spectrum of the nanoparticle dressed by lasers of varying amplitude  $\varepsilon_d \in [0, 0.4~ \text{V/\AA}]$. Blue lines refer to stimulated emission, red lines to net absorption and the grey lines signal the absorption peaks. Convergence of the absorption spectra requires considering $N_F=61$ Fourier components symmetrically around $n=0$.  At equilibrium ($\varepsilon_d=0.00$), the nanoparticle is transparent in the 0-3.2 eV range. Optical transitions start to appear when the frequency of the probing pulse is larger than the band-gap. The absorption spectra completely changes as the system is driven far from equilibrium even when the driving pulse is not generating any net charge carriers in the conduction band. 

There are three essential features that emerge in the absorption as the electronic system is driven out of equilibrium: (i) \emph{Below bandgap absorption.} As the driving amplitude increases in the  $0-0.20$ V/\AA~range we observe the emergence of additional spectral features just below the 3.2 eV bandgap. This phenomenon has been experimentally  observed before \cite{Ghimire2011, Kono2004} and is reminiscent of the dynamic Franz-Keldysh effect~\cite{Keldysh1958} in solids and the quantum confined Stark effect in nanostructures \cite{Miller1984}.  (ii) \emph{Broadband absorption.} A novel feature that is predicted by the theory is that by driving the nanoparticle non-resonantly it is possible to reversibly turn this IR/Vis transparent material into a broadband absorber. For instance, the non-equilibrium spectra for $\varepsilon_d=0.24-0.36$ V/\AA~ exhibits several novel absorption peaks across the IR/Vis region that are spaced by the photon energy of  the  driving  light.  (iii) \emph{Low-frequency absorption/stimulated emission.} Another novel feature that emerges far from equilibrium are strong absorption and stimulated emission features in the THz region of the electromagnetic spectrum, as those exhibited for $\varepsilon_d=0.20$ V/\AA (emission) and $\varepsilon_d=0.26$ V/\AA (absorption). By  driving the system out of equilibrium by non-resonant light it is possible to completely change the absorption spectra of the driven materials in a reversible fashion and tune its optical properties. We now interpret these three basic features in the non-equilibrium absorption from a Floquet perspective. 

\subsubsection{Below bandgap absorption}

\begin{figure*}[htbp]
	\centering
	\includegraphics[width=0.7\textwidth]{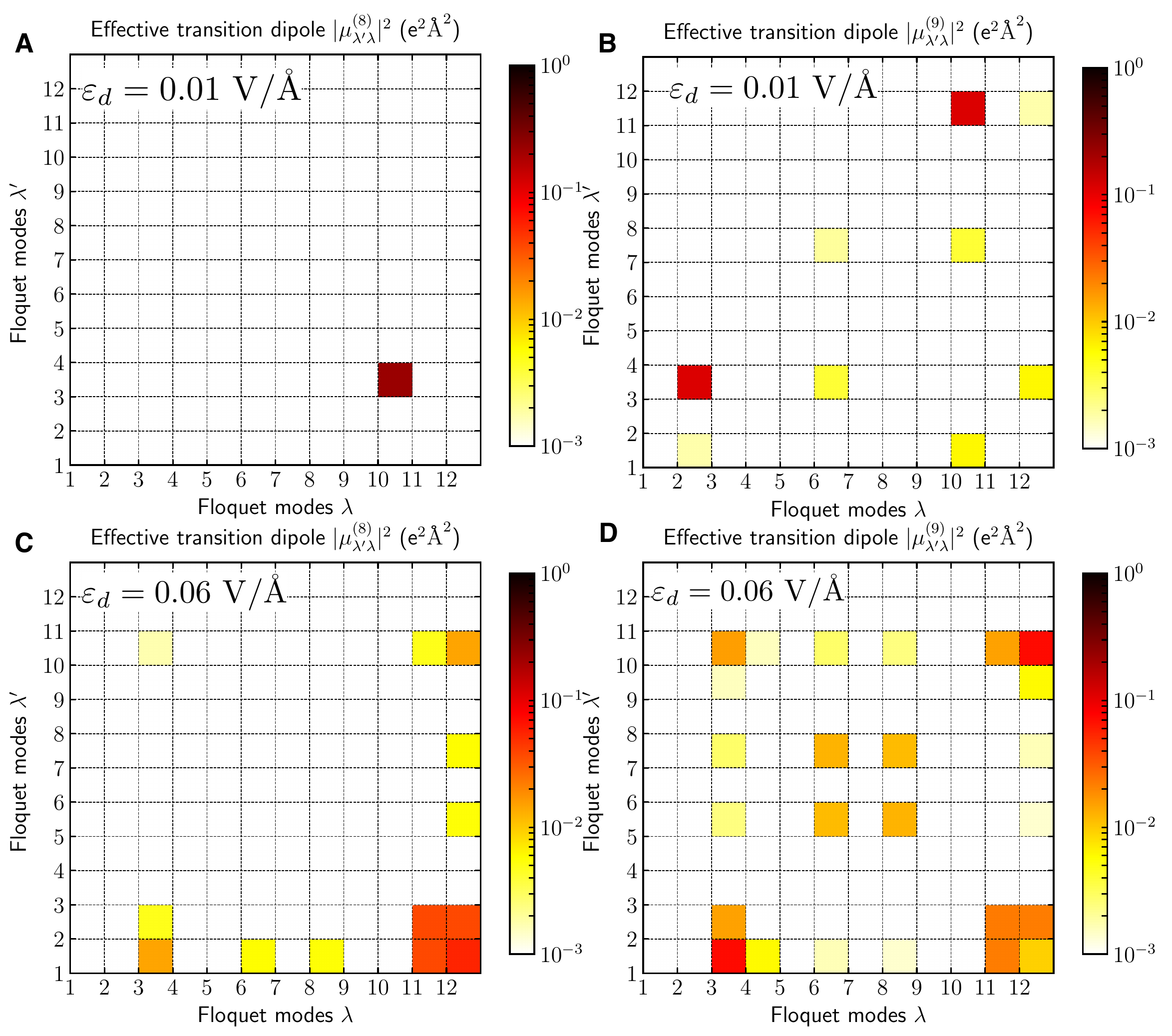}
	\caption{ Inter-BZ transition dipoles $\mu_{\lambda', \lambda}^{(n)}$ between Floquet modes with indexes $\lambda$ and $\lambda'$ separated by $n$ Brillouin zones. The figure contrasts results for weak $\varepsilon_d = 0.01$ V/\AA\ (A,B) and stronger $\varepsilon_d = 0.06$ V/\AA~ (C,D) driving field amplitudes. Note how the transition amplitudes for $n=8$ and $n=9$ for $\varepsilon_d = 0.06$ V/\AA~are approximate replicas of one another.  }
	\label{fig:dipole_comp}
\end{figure*}

As the driving field amplitude is increased up to 0.06 V/\AA~the first thing that is observed is the emergence of an additional series of absorption peaks around 3.0-3.2 eV and a reduction of the intensity of the peaks around 3.3-3.4 eV. These new peaks appear less than $\hbar\Omega$ away from the main absorption features at equilibrium and lead to a net red shift in the absorption spectrum.  This phenomenon can be understood in the context of \eq{eq:final3} by examining the transition dipoles of the driven system. As shown in \fig{fig:dipole_comp}, as the driving field is increased from 0.01 V/\AA~(A-B) to 0.06 V/\AA~(C-D) there is an increase  in the magnitude of the transition dipoles between Floquet modes that are $n=8$ BZs away that are responsible for these new spectral features. From \fig{fig:dipole_comp} it is also clear that for $\varepsilon_d = 0.06$  V/\AA, $\mu_{\lambda', \lambda}^{(8)}$ and $\mu_{\lambda', \lambda}^{(9)}$ are approximate replicas of one another. This indicates that these new spectral features can be understood as Floquet replicas of the transitions for \emph{the non-equilibrium material}. By contrast when the system is close to equilibrium (i.e. for $\varepsilon_d = 0.01$  V/\AA), $\mu_{\lambda', \lambda}^{(8)}$ and $\mu_{\lambda', \lambda}^{(9)}$ are completely different and just reflect the transition dipoles of the pristine material. 

\subsubsection{Broadband absorption and spectral replicas}

\begin{figure*}[htbp]
	\centering
	\includegraphics[width=0.7\textwidth]{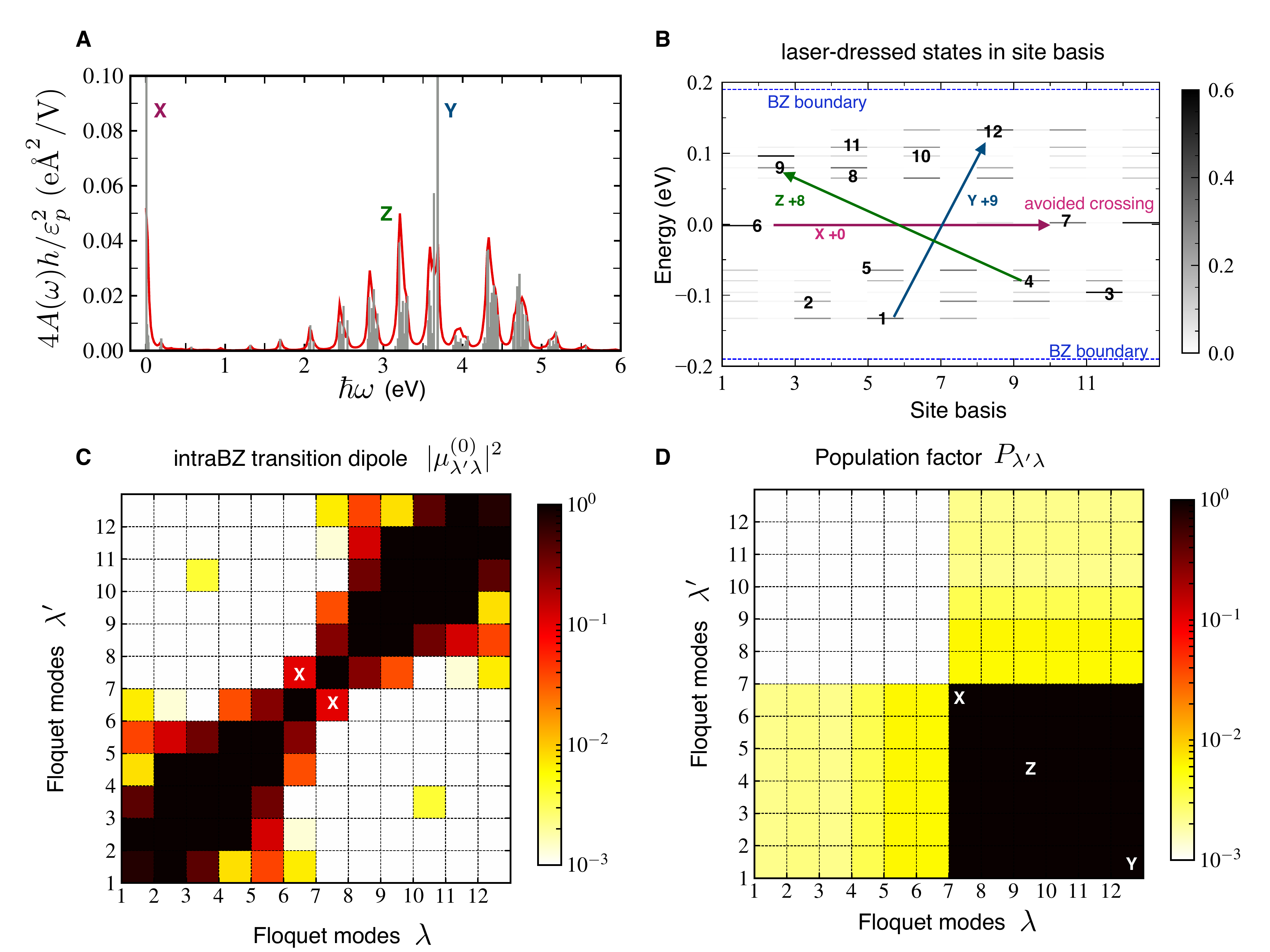}
	\caption{Interpretation of the absorption spectrum for $\varepsilon_d = 0.26$ V/\AA. The absorption spectrum (A) of the laser-dressed material exhibits a strong absorption feature at low frequency (labeled X), and several absorption features in the IR/Vis/UV region that are periodically separated by $\hbar\Omega$ which make the material a broadband absorber. Two of them are labeled Y and Z.  (B) Quasienergies of the Floquet modes in the first BZ and population $|\braket{k|\phi_{\lambda}^0}|^2$ of the Floquet modes in the site basis at times $t_0+nT$. The arrows signal optical transitions responsible for the X, Y, Z spectral features in (A), and the $+n$ the number of BZs separating the two Floquet modes involved in the transition,  e.g., X +0 means this is an intraBZ transition. (C) Effective dipole between Floquet states within the first BZ  $|\mu^{(0)}_{\lambda \lambda'}|^2$. (D) Effective population factor $P_{\lambda \lambda'}$ ($\lambda'\to\lambda$) between Floquet modes. The numeric labels to the Floquet modes in (C-D) are those assigned in (B). } 
	\label{fig:E0.26} 
\end{figure*}

 As we increase the amplitude of the driving laser to $\varepsilon_d = 0.26$ V/\AA~ (\fig{fig:ZnO}), the semiconducting material becomes a broadband absorber with the emergence of new absorption features in the IR/Vis region where the material was transparent and with a change in the spectral features in the UV region of the pristine material. Interestingly, the absorption spectra exhibits a clear periodic structure with spectral features separated by multiples of the driving photon energy $\hbar\Omega$. 
 
 To understand these features consider \fig{fig:E0.26} which details the properties of the Floquet modes, population factors and the intraBZ transition dipoles. The quasienergies and distribution of the Floquet modes along the chain is shown in \fig{fig:E0.26}B.  Floquet states 1-6 have quasienergies $0>\mc{E}_\lambda>-\hbar \Omega/2$ and are composed mostly of Wannier states that form the valence band of the pristine material (of odd site states). In turn, Floquet states 7-12 have quasienergies $0<\mc{E}_\lambda < \hbar \Omega/2$ and are composed mostly of Wannier states that form the conduction band. In this case, the population factors allow valence to conduction band transitions that originate in $\lambda'=1-6$ and end in $\lambda=7-12$. We observe that the Floquet modes are delocalized across the nanoparticle (\fig{fig:E0.26}B).  Interestingly, the degree of delocalization of the Floquet modes is smaller than the states of the pristine system, but larger than the Wannier-Stark states \cite{Wannier1960, Holthaus1996} that would have been obtained by diagonalizing the Hamiltonian in the presence of a fixed electric field with amplitude 0.26 V/\AA. 

We label the largest absorption feature in the two main cluster of peaks around 3.2 and 3.6 eV in the absorption spectra (\fig{fig:E0.26}A) by Z and Y, respectively.  As can be seen in \fig{fig:E0.26}B, Y originates from transitions between  Floquet modes $\ket{\phi_1} \rightarrow \ket{\phi_{12}}$ separated by $n = 9$ BZs, while $Z$ originates from  $\ket{\phi_4} \rightarrow \ket{\phi_9}$ transitions separated by $n=8$ BZs. Thus, the cluster of peaks separated by the driving photon energy originate from transitions that are separated by a different number of Floquet replicas.  The absorption spectra has visible transitions between BZs that are separated by $n$=3-14 BZs. Generally, the magnitude of the absorption decreases as the $n$ BZs that separate a given transition deviates from the $n=8-9$ needed for a transition across the band gap in the near-equilibrium system.  The contributions coming from lower $n$ broaden the frequency regime for the absorption of the material, and turn it into a broadband absorber. 
 
The periodic structure in the absortion spectra is a clear manifestation of the periodicity of the Floquet space. This remarkable feature is particularly evident for some special values of the driving electric field (see, e.g. 0.28 V/\AA~in \fig{fig:ZnO}). For such values, 10 of the 12 Floquet states cluster around 2 particular quasienergy values in the first BZ, leading to a spectra with sharp periodic features. The remaining two Floquet states remain close to 0 and their relevance is discussed in \stn{sec:lowfreq}. This  spectral signature of the Floquet modes is complementary to those in photoemission spectroscopy \cite{Wang2013}.

\subsubsection{Low-frequency spectral features}
\label{sec:lowfreq}

Surprisingly, for particular values of the driving electric field we observe a strong low-frequency ($\sim$meV) absorption or stimulated emission band, see for example $\varepsilon_d = 0.20, 0.26, 0.36$ V/\AA. These  novel features can be probed using THz radiation or be used to generated THz pulses. To understand the underlying physics from a Floquet perspective, consider the transition that leads to this phenomenon for $\varepsilon_d = 0.26$ V/\AA~ labeled as X in \fig{fig:E0.26}A. As shown in \fig{fig:E0.26}B, we identify the strongest low-frequency transition at $\hbar\omega = 4$ meV  as the intraBZ transition from Floquet modes 6 to 7. These two states are both dipole and population allowed, see \fig{fig:E0.26}C-D. 

\begin{figure*}[htbp]
	\centering
	\includegraphics[width=0.8\textwidth]{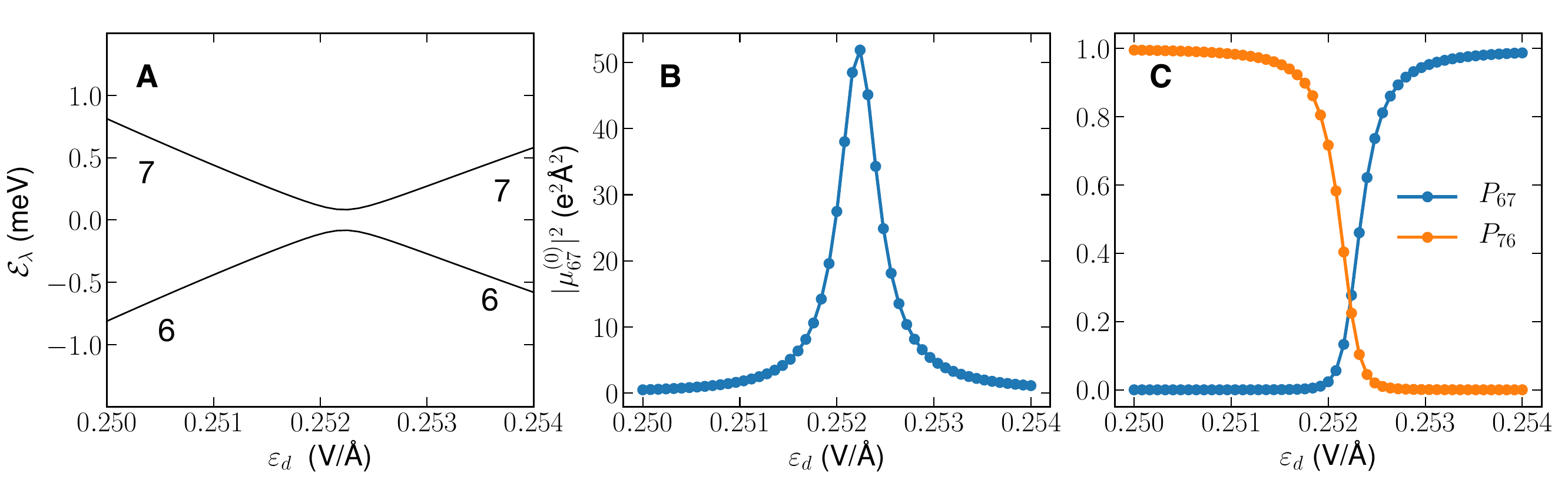}
	\caption{(A) Quasienergy, (B) intraBZ transition dipole and (C) population factors around the  avoided crossing occuring at  $\varepsilon_{d, \text{crossing}}= 0.2522$ V/AA~ in the first BZ between Floquet modes 6 and 7.  The hybridization of the two Floquet modes around the crossing leads to large intraBZ effective transition dipole $|\mu_{67}^{(0)}|^2$, population mixing between the two states, and absorption/stimulated emission features at meV frequencies. }
	\label{fig:avoided}
\end{figure*}

\begin{figure}[htbp]
	\centering
	\includegraphics[width=0.4\textwidth]{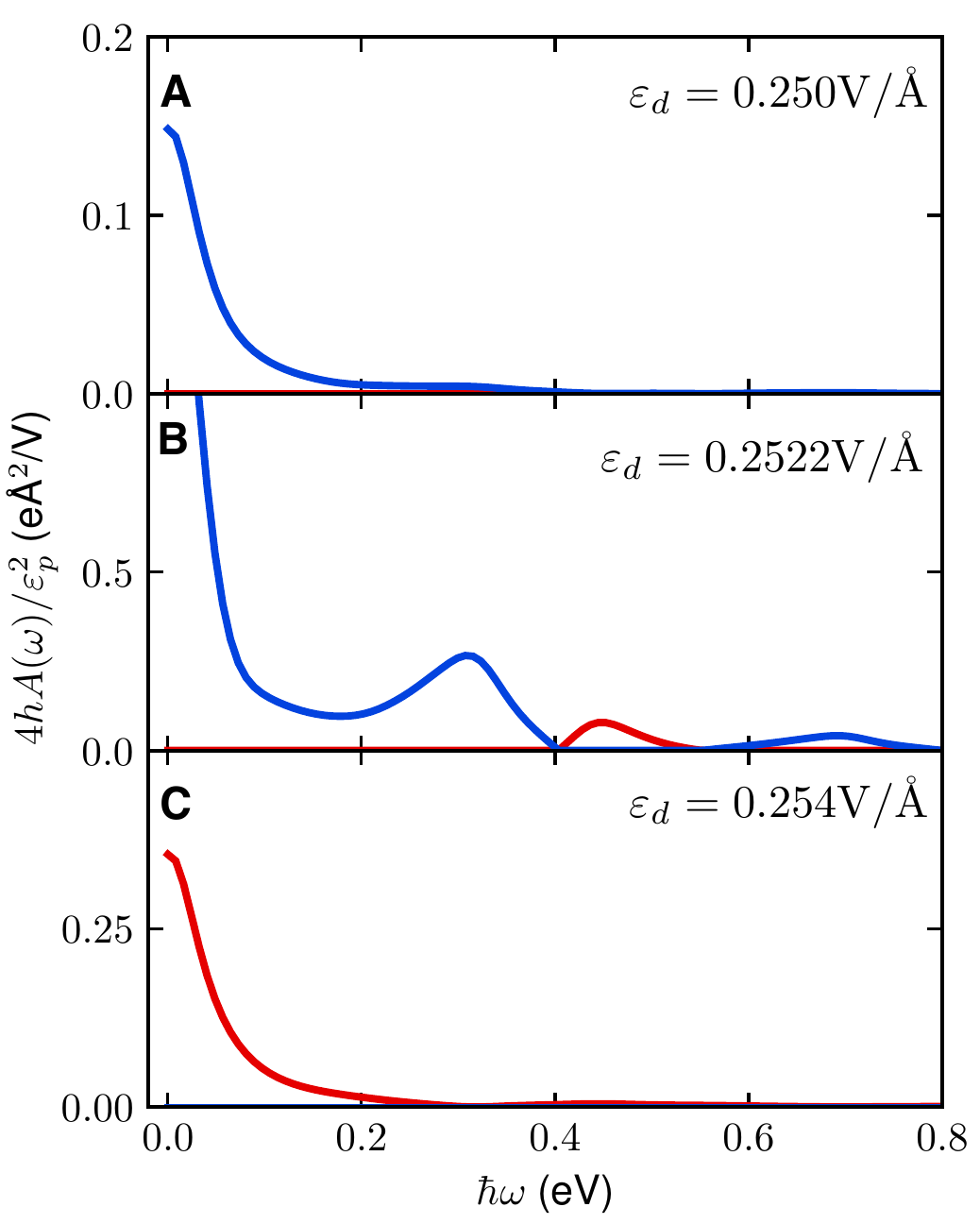}
	\caption{ Net absorption (red) or stimulated emission (blue) spectra at low frequencies for a driving pulse (A)  just below  the avoided crossing shown in \fig{fig:avoided}A, (B) right at the crossing  $\varepsilon_{d}=\varepsilon_{d, \text{crossing}}$ and (C) just above  it.  The low frequency absorption spectra transitions from stimulated emission to absorption as the driving electric field amplitude is increased across the avoided crossing.}
	\label{fig:absorption_avoided} 
\end{figure}

As shown in \fig{fig:avoided}A, the Floquet modes 6 and 7 form an avoided crossing in the Floquet picture as the driving amplitude is changed around $\varepsilon_{d, \text{crossing}}= 0.2522$ V/\AA. Away from the avoided crossing these two states do not have a significant intraBZ transition dipole.  However, as shown in \fig{fig:avoided}B, the hybridization of the two Floquet modes around the avoided crossing creates a strong transition dipole between the two levels that peaks at the crossing point $\varepsilon_{d, \text{crossing}}$. Such hybridization leads to very large absorption and stimulated emission features in the absorption spectra at low frequencies. In fact, as shown in \fig{fig:absorption_avoided}, these low frequency transitions are an order of magnitude stronger than even the largest  absorption peak observed at equilibrium. The hybridization also open a small energy gap that imposes a lower limit to the frequency of the transition that can be observed, in this case $\sim 0.2$ meV.  

Around the crossing, both absorption and stimulated emission are present. The dominant phenomenon depends on the population factor as the absolute value of the effective dipole is the same for both transitions (i.e., $|\mu_{\lambda'\lambda}^{(0)}| = |\mu_{\lambda \lambda'}^{(0)}|$). As shown in \fig{fig:absorption_avoided},  stimulated emission dominates for driving electric fields $\varepsilon_{d}< \varepsilon_{d, \text{crossing}}$ because in this case $P_{76} > P_{67}$ (\fig{fig:avoided}C), while absorption dominates for $\varepsilon_{d}> \varepsilon_{d, \text{crossing}}$ because $P_{67} > P_{76}$.  For a driving electric field with amplitude right around the avoided crossing (\fig{fig:absorption_avoided}B), a rich spectrum with both absortion and stimulated absorption features results.  The additional peaks around $\hbar\omega$ = 0.38 eV originate from transitions from Floquet modes 6 and 7 to equivalent states one BZ away.

 The avoided crossings in the Floquet picture is responsible for a number of novel phenomena such as bond softening and hardening of diatomic molecules under intense laser driving \cite{Frasinski1999, Sheehy1996} and coherent destruction of tunneling \cite{Grossmann1991} under high-frequency driving. In this context it leads to strong low frequency absorption and emission due to hybridization of Floquet states. 
 
\section{Discussion} \label{sec:conclude}

\subsection{Summary of  observations}

In summary, we have developed a general theory to study the optical absorption properties of laser-driven materials. 
The optical absorption in this non-equilibrium case is defined as the rate of transitions between laser-dressed states due to interaction with the probe laser [\eq{eq:definition}]. By treating the probe laser in first-order perturbation theory, it is  possible to relate the non-equilibrium absorption spectra to the two-time dipole-dipole correlation function in the interaction picture of the laser-dressed Hamiltonian [\eq{eq:def4}]. To make further progress we focused on effective non-interacting electronic systems for which the dynamics of the creation and annihilation operators in the driving pulse can be solved in closed form by invoking Floquet theorem [\eq{eq:sol}].  In this way, we were able to treat the interaction between the driving electric field and matter exactly and reduce the complex time-dependent non-equilibrium calculations to a time-independent diagonalization in an extended Hilbert space.  

These developments lead to a final expression [\eq{eq:final3}] for non-equilibrium optical absorption which has a similar structure to the equilibrium one. In it, the Floquet modes play the role of system eigenstates and there are contributions due to absorption and stimulated emission. Transitions are allowed when a probe photon is at resonance with the transition frequency between two Floquet modes that have a non-zero transition dipole and that are allowed by populations. While the investigation of laser-matter interactions using Floquet approaches usually focuses on resonances between Floquet states \cite{Frasinski1999, Grossmann1991,Fleischhauer2005}, in this theory the focus is on the optical transitions induced by the probe light between Floquet modes. One unique feature of the non-equilibrium absorption theory is that the transition dipoles carry an additional index indicating the number of Brillouin  Zones separating the two Floquet modes. 

To test the validity of the theory, we employed it to recover and interpret the well-known Autler-Townes effect. We further used the theory to characterize the non-equilibrium absorption of a model semiconducting nanoparticle reversibly driven far from equilibrium by  non-resonant light. The computational analysis recovered the previously observed below band gap absorption \cite{Ghimire2011, Kono2004} and revealed two new phenomena: (i) Nonresonant light turns this IR/Vis transparent material into a broadband absorber with multiple absorption features in the energy gap of the pristine material. These features are periodically spaced by the driving photon energy and are a characteristic signature of the periodic structure of Floquet space. They  can be used as an optical signature of the presence of Floquet states.  (ii) Non-resonant light opens strong  low frequency ($\sim$ meV) absorption and stimulated emission features  at particular driving amplitudes. These features arise because of transitions between nearly degenerate Floquet modes  that hybridize thus enhancing their transition dipole. Such pair of states observe an avoided crossing with increasing driving electric field amplitude.  Both  low frequency absorption or stimulated emission can be observed and tuned by changing the driving electric field amplitude around the avoided crossing. These three significant changes in the absorption properties of the model nanoparticle are present in a reversible regime of the laser-matter interaction where the driving pulse \emph{per se} does not generate real carriers.

\subsection{Qualitative picture of non-equilibrium absorption}

\begin{figure*}[htbp]
	\centering
	\includegraphics[width=0.75\textwidth]{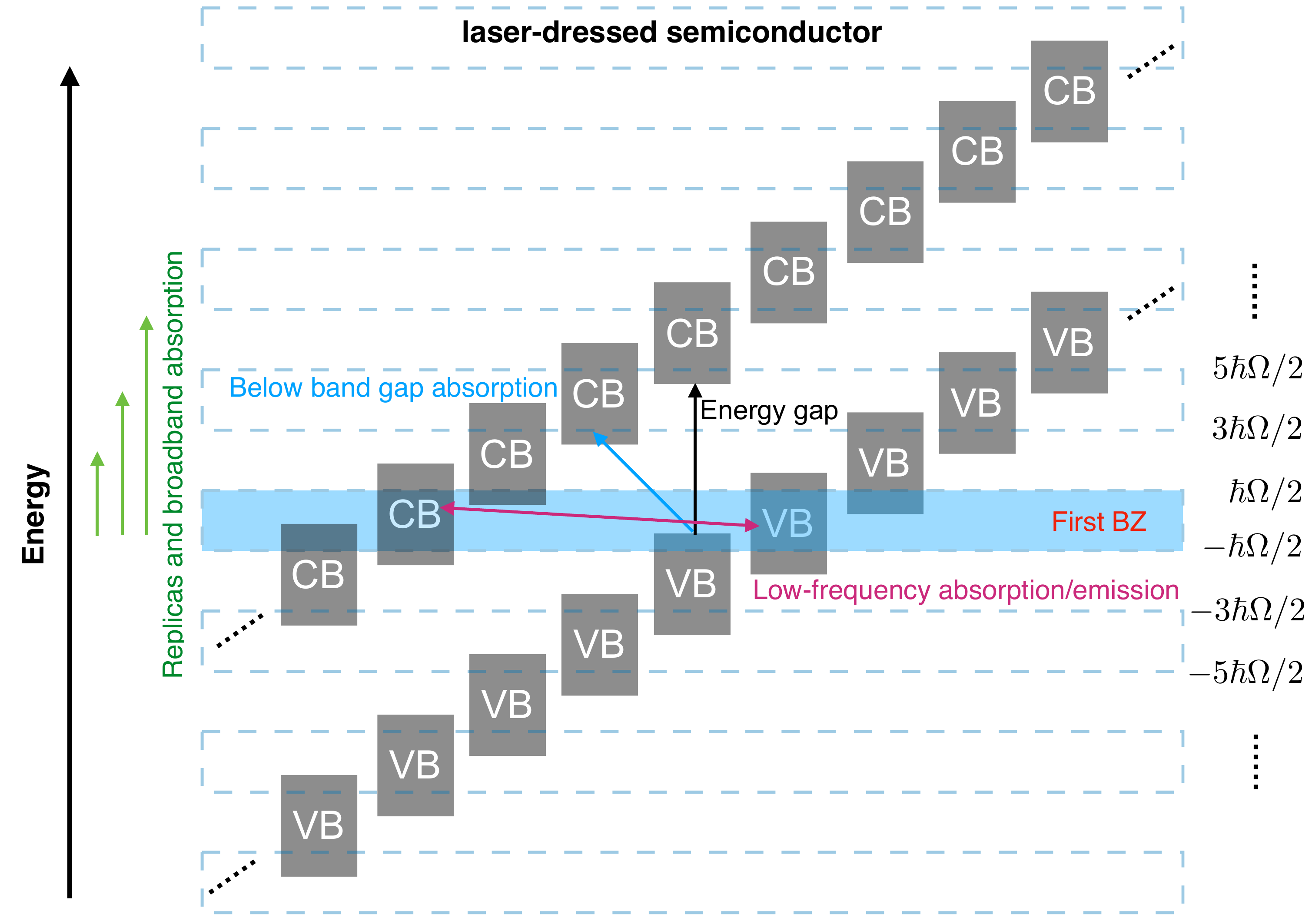}
	\caption{ Schematic of the energy diagram of a semiconductor dressed by  non-resonant light of frequency $\Omega$. The driving creates Floquet replicas of the valence (VB) and conduction (CB) band levels of the pristine material separated by integer $n\hbar \Omega$, leading to novel features in the absorption spectrum including below band gap absorption, low frequency transitions and broadband absorption. }
	\label{fig:nonresonant} 
\end{figure*}

At this point, it is useful to summarize  these observations into a  qualitative picture of the non-equilibrium optical absorption spectra.  \Fig{fig:nonresonant} shows a schematic energy diagram of a semiconducting system in the presence of non-resonant driving light. In a Floquet sense, the dressing by the driving pulse leads to replicas of the valence and conduction band of the material that are separated by multiples of the driving pulse photon energy $n\hbar\Omega$. When the driving laser is weak, only optical transitions across the band gap $E_g$ of the material are allowed. The levels involved are separated at least $m=\floor{E_g/\hbar\Omega}$  BZs away, where $\floor{\cdot}$ denotes the floor function. For simplicity in presentation, let us suppose that $E_g$ is precisely $m$ BZs away.  Thus, the transition dipoles $\mu_{\lambda, \lambda'}^{(n)}$ for $n< m$ will be zero. As the amplitude of the driving laser field $\varepsilon_d$ is increased the replicas are distorted and new, previously forbidden, interBZ transition dipoles are created. The below band gap absorption occurs when those at $n=m-1$ are allowed and these features are separated from $E_g$ at most by $\hbar\Omega$. As $\varepsilon_d$ is increased additional absorption features are created for $n=m-2, m-3, m-4, \cdots$ leading to periodic absorption features that are separated by multiples of $\hbar\Omega$ and that make this initially transparent semiconductor into a broadband absorber. Low frequency ($\sim$ meV) optical features emerge when there are optically accessible intraBZ transitions or transitions between Floquet modes at adjacent BZ edges, as schematically shown in the figure. For non-resonant driving most of these transitions will be optically forbidden either through transition dipoles or population factors. Strong low frequency transitions are opened when a pair of Floquet modes that are allowed by population factors enter into resonance and through hybridization create strong intraBZ transition dipoles.

\subsection{Floquet prospects}

 \Eq{eq:final3} indicates that the natural states to interpret the nonequilibrium absorption spectra are the Floquet modes and not the pristine states of the system. In terms of interpretation, this transition from pristine states to Floquet modes   leads to a series of important changes in our intuition:  (i) The transition dipoles between Floquet modes are time-dependent, periodic in time and admit a  Fourier expansion  [\eq{eq:mufourier}]  with component $\mu_{\lambda\lambda'}^{(n)}$ oscillating at frequency $n\Omega$. The $\mu_{\lambda\lambda'}^{(n)}$ component connects Floquet states that are $n$ BZs away, and determine which transitions open as the driving laser changes. This contrasts with the equilibrium case in which  the transition dipoles between pristine eigenstates are time-independent quantities. (ii) Because the Floquet modes are time-dependent, to develop intuition into which modes are dipole connected it is important to examine their spatial distribution at all times $t_0<t\le t_0+T$ and not just at a particular time $t_0$. (iii) The population factors that determine which transitions are allowed depend on the initial state \emph{and} the driving light.  This contrasts with the equilibrium theory of absorption where this is just determined by the initial distribution of population among states. (iv) A natural subdivision of energy is the driving photon energy $\hbar\Omega$ that separates the different BZs. As schematically shown in \fig{fig:nonresonant}, one can interpret the optical properties by focusing on a given BZ and examining inter and intraBZ transitions.  Except for high frequency driving, the relevant Floquet transitions that signal interband absorption will not be for states in the same BZ but for those separated $\floor{E_g/\hbar\Omega}$ Floquet BZs away.

The developed theory and interpretation scheme can be used to design laser-driven materials with desirable non-equilibrium optical properties. It applies for effective non-interacting electronic materials driven resonantly or non-resonantly by multichromatic light with commensurate frequencies of arbitrary strength. Future prospects include computations in realistic materials using \emph{ab initio} based models, extending the theory to the fully quantum regime, and characterizing the role of electron-electron and electron-phonon interactions in the non-equilibrium absorption. 

\begin{acknowledgments}
	This material is based upon work supported by the National
	Science Foundation under CHE-1553939.	
\end{acknowledgments}

\appendix

\section{Nullity of the neglected term in Eq. (43)}
The neglected term in Eq. (43) is 
\be
I(\omega) =   \frac{|\varepsilon_p|^2}{4\hbar^2}  \lim_{t \rightarrow +\infty} \frac{1}{ t - t_0}\iint_{-\infty}^{t} C_{\mu \mu}(\bar{t},\tau) (e^{-i 2\omega \bar{t}} + \text{c.c.})\, d\bar{t}d\tau \\ 
\label{eq:tmp10}
\ee
where 
\be 
\begin{split}
C_{\mu\mu}(\bar{t},\tau)  = & \sum_{n,n'}\sum_{\lambda, \lambda',\eta,\eta'} \sum_{\gamma \delta \gamma' \delta'} D^n_{ \lambda' \lambda \gamma \delta}  D^{n'}_{\eta' \eta \gamma' \delta'} \\ 
&\times  e^{i (\mc{E}_{\eta'\eta}  + \mc{E}_{\lambda'\lambda})(\bar{t}-t_0)/\hbar + i(n'+n) \Omega  \bar{t} } \nonumber \\ 
& \times e^{i ((\mc{E}_{\eta'\eta} - \mc{E}_{\lambda'\lambda})/\hbar + (n'-n) \Omega) \tau/2} \braket{ c^\dagger_\gamma  c_\delta  c^\dagger_{\gamma'}  c_{\delta'} } \label{eq:corr}. 
\end{split} 
\ee 
To make an appreciable contribution,  the argument in exponent needs to vanish in order to cancel the factor $1/(t-t_0)$. Consider the $e^{-i2\omega \bar{t}}$ term in \eq{eq:tmp10} (the other term can be obtained simply by $\omega \leftrightarrow -\omega$). For the argument to vanish,
\be   \mc{E}_{\eta'\eta} + \mc{E}_{\lambda'\lambda} = 0 , ~~~  (n'+n)\Omega - 2\omega = 0 . 
\label{eq:tmp11}
\ee 
In turn, the integration with respect to  $\tau$ yields the following delta function 
\be  \delta( (\mc{E}_{\eta'\eta} - \mc{E}_{\lambda'\lambda} + (n'-n)\hbar\Omega)/2)  
\ee 
Inserting the first equality in \eq{eq:tmp11} into this delta function yields $\delta(\mc{E}_{\eta'\eta} - (n-n')\hbar\Omega/2)$. Thus, this term is non-zero when $\Omega = 2\mc{E}_{\eta'\eta}/\hbar(n-n')$. This condition, in general, is not satisfied except accidentally. Thus the neglected term does not contribute to the final absorption spectrum. 


%

\end{document}